\documentclass[a4paper,11pt]{article}
\pdfoutput=1 % if your are submitting a pdflatex (i.e. if you have
\usepackage{jheppub} % for details on the use of the package, please
\usepackage[T1]{fontenc} % if needed
\usepackage{braket}
\newcommand{\be}{\begin{equation}}
\newcommand{\ee}{\end{equation}}
\newcommand{\bea}{\begin{eqnarray}}
\newcommand{\eea}{\end{eqnarray}}

\def\be{\begin{equation}}
\def\ee{\end{equation}}
\def\ba{\begin{eqnarray}}
\def\ea{\end{eqnarray}}

 \def\ba{{\bar{\alpha}}}

\definecolor{Gray}{gray}{0.95}
\definecolor{LightCyan}{rgb}{0.88,1,1}

\title{Conformal field theory and the web of quantum chaos diagnostics}

\author[a]{Jonah Kudler-Flam}
\author[a]{Laimei Nie}
\author[a,b]{Shinsei Ryu}

\affiliation[a]{Kadanoff Center for Theoretical Physics, University of Chicago, Chicago, IL~60637, USA}
\affiliation[b]{James Franck Institute, University of Chicago, Chicago, Illinois 60637, USA}

\emailAdd{jkudlerflam@uchicago.edu}
\emailAdd{nlm@uchicago.edu}
\emailAdd{ryuu@uchicago.edu}

\abstract{We study three prominent diagnostics of chaos and scrambling in the context of two-dimensional conformal field theory: the spectral form factor, out-of-time-ordered correlators, and unitary operator entanglement. With the observation that all three quantities may be obtained by different analytic continuations of the torus partition function, we address the connections and distinctions between the information that each quantity provides us. In this process, we study the emergence of irrationality from ``large-N" limits of rational conformal field theories (RCFTs) as well as the explicit breakdown of rationality for theories with central charges greater than the number of their conserved currents. Our analysis begins to elucidate the intermediate dynamical behavior of theories that bridge the gap between integrable RCFTs and maximally chaotic holographic CFTs.}

\begin{document} 
\maketitle
\flushbottom

\section{Introduction \& background}
\label{sec: intro}

\begin{figure}
    \begin{center}
         \includegraphics[height=7cm]{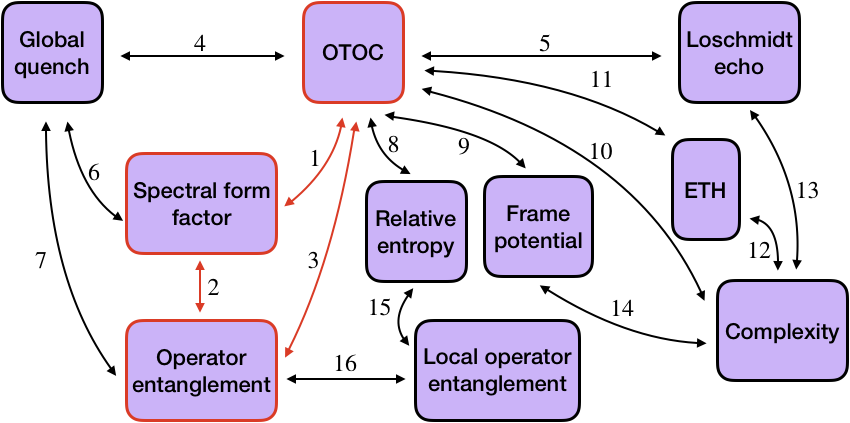}
    \end{center}
    \caption{We lay out a web of chaos diagnostics. In this paper, we focus on the three highlighted in red. The numbered connections have been studied as follows: (1-3) Related in this paper through analytic continuations of the torus partition function. (1) Higher point OTOCs were related to higher-point spectral form factors in Ref.~\cite{2017JHEP...11..048C} and averaged OTOCs were further equated with the SFF in quantum field theory in Refs.~\cite{2019PhLB..795..183D,2019arXiv190704289M}. (3) In spin systems, the R\'enyi operator mutual information is also directly related to the average OTOC \cite{2016JHEP...02..004H}. (4,6-7) R\'enyi entropy of disjoint intervals after a global quantum quench may also be computed by a particular analytic continuation of the torus partition function \cite{2015JHEP...09..110A}, though we do not focus on this in this paper. (5) The OTOC was equated with the thermally averaged Loschmidt echo in Refs.~\cite{2019arXiv190302651Y,2019JHEP...07..107R}. This was further explored in Ref.~\cite{2019arXiv190901894B}. (8) The relative entropy of perturbed thermal states was related to OTOC in Ref.~\cite{2018JHEP...07..002N}. (9) $2k$-point OTOCs were equated with $k^{th}$ frame potentials in Ref.~\cite{2017JHEP...04..121R}. (10-12) The rate of growth of Lanczos coefficients, a notion of operator complexity was shown to bound the Lyapunov exponent of OTOCs in Ref.~\cite{2018arXiv181208657P}. Both the bounds on this rate and the Lyapunov exponent were shown to follow from the eigenstate thermalization hypothesis in Ref.~\cite{2019arXiv190610808M}. (13) The circuit complexity was suggested to be related to the logarithm of the Loschmidt echo in Ref.~\cite{2019arXiv190901894B}. (14) The frame potentials were shown to bound the circuit complexity in Ref.~\cite{2017JHEP...11..048C}. (15-16) The entanglement content of a Heisenberg time-evolved local operator may be shown to be related to relative entropy of excited states and the OTOC of the local operator with twist fields \cite{2020arXiv200514243K}}
    \label{fig:cartoon}
\end{figure}

Many-body quantum chaos has garnered immense attention in recent years from a variety of fields due to its key role in the understanding of the emergence of thermal physics \cite{PhysRevA.43.2046,1994PhRvE..50..888S} and the emergence of Einstein gravity \cite{2014JHEP...03..067S,2015PhRvL.115m1603R,2016JHEP...08..106M}. Given this attention and the inherent complexity of characterizing quantum chaos, a slew of diagnostics have been proposed, though there is no general consensus on which ones are superior. %Furthermore, 
Fig.~\ref{fig:cartoon} illustrates the web of various chaos diagnostics\footnote{We have taken this terminology from Ref.~\cite{2019arXiv190901894B}.} that have been studied in the literature. Previous efforts from high energy, condensed matter, and quantum information communities have partially established the connections among different measures in selective models, but a more general and unifying scheme remains elusive. Moreover, most of these results have been in the context of systems with finite dimensional local Hilbert spaces. One of our main goals is to extend these connections to quantum field theory.

In this work, we begin to address this question by connecting three prominent diagnostics in the web: the spectral form factor (SFF), out-of-time-ordered correlators (OTOC), and operator entanglement. While each one of these initially seems like a very unique diagnostic, we show that in the context of 2D conformal field theory (CFT), all three may be computed by different analytic continuations of the torus partition function. As such, the three aforementioned quantities are three different ways of reading out the information that the torus partition function contains regarding the integrability of the theory. It is natural that the partition function becomes the object of central interest in that it is directly related to the spectrum of the theory, the statistics of which are often considered a smoking gun for chaos\footnote{Generally, the level spacings in integrable theories obey Poisson statistics, $e^{-\omega}$, while non-integrable theories follow the Wigner surmise, $\frac{\pi \omega}{2}e^{-\pi \omega^2/4}$ \cite{2016AdPhy..65..239D}.}.

In the rest of the introduction, we introduce the three main players and demonstrate how they unify in conformal field theory.

\paragraph{Spectral form factor} 
Spectral features of a Hamiltonian are good indicators of the ergodic nature of a system. While the level spacing of a Hamiltonian is certainly revealing, the spectral form factor contains further information because it probes the level statistics of both close and far-separated eigenvalues. The spectral form factor is defined as
\begin{align}
    g(\beta, t) \equiv \left|\mathcal{Z}(\beta+ it) \right|^2 =  \sum_{n,m} e^{-\beta(E_m + E_n)} e^{i(E_m - E_n)t}
\end{align}
where the inverse temperature $\beta$ is present as a regulator to cut off the high-energy modes. 
Later ``times" probe correlations between eigenvalues that are closer.
For systems with discrete spectra, the SFF plateaus in the late time limit, where the time scale at which the plateau occurs is determined by the minimal level spacing. This plateau is due to only the $m=n$ terms contributing significantly to the sum because the others are oscillating wildly. Assuming no degeneracy in the spectrum, the plateau magnitude will be $\mathcal{Z}(2\beta)$.
In comparison, during the earlier time range, the corresponding energy scale that is being probed is much larger than the mean nearest level spacing, and the SFF becomes sensitive to the ``spectral rigidity," namely the repulsion of energy levels that are far away from each other. In the presence of such rigidity, the SFF displays a ramp feature that increases linearly with time before transitioning into the plateau.
Before the spectral rigidity sets in, the SFF typically exhibits a non-universal decrease in time, which is expected from the increasing cancellations from the oscillatory phases as $t$ becomes non-zero. This initial reduction together with the ramp leads to a dip feature in the SFF.

The prominent dip-ramp-plateau structure of the SFF has been long studied in the context of random matrix theory, where the detailed shape of the (ensemble averaged) curve depends on the particular ensemble that the matrices are drawn from. It has also been observed in several chaotic models, such as the SYK model~\cite{2017JHEP...05..118C}, with the shape determined by the symmetries of the Hamiltonian. Generically, one has to employ an averaging scheme to find the characteristic dip-ramp-plateau. While there is a natural averaging scheme when drawing coupling constants from an ensemble as in SYK, the averaging is more subtle in theories with fixed Hamiltonians.
The early-time decay of the spectral form factor is self-averaging, whereas the late-time oscillations can be controlled by a ``progressive time averaging" scheme\footnote{The conclusions we arrive at in this paper are largely insensitive to the averaging scheme. However, we empirically find the progressive time averaging to capture the coarse-grained features of the spectral form factor most cleanly. Furthermore, this averaging scheme does not require one to have an ensemble of theories.}~\cite{2017JHEP...03..154B}
\be
g_{\mbox{\tiny prog}}(\beta, t) = \frac{1}{100} \sum\limits_{k = -50}^{50} g(\beta, t+\frac{k}{100}  \alpha  t)
\ee
where the linear-$t$ time window for averaging is equivalent to fixed time window in $\log t$, $0 < \alpha < 2$ is a constant that can be tuned to minimize deviations from the early self-averaging part of the curve, and the number of time steps 100 in each time window is chosen empirically to ensure the smoothness of the curve at late times and to be computationally tractable.
As a convenient alternative averaging scheme, progressive time averaging of a single realization of random matrix was shown to provide a good approximation to the usual ensemble average for the Gaussian random matrices~\cite{2017JHEP...03..154B}, in contrast to ordinary time averaging with fixed time window which either significantly deviates for the early-time value or fails to effectively suppress the late-time noise.

\paragraph{OTOC}
The out-of-time-ordered correlator (OTOC) \cite{1969JETP...28.1200L} has been the central object studied when searching for a diagnostic of the butterfly effect in quantum systems~\cite{2014JHEP...03..067S, 2015PhRvL.115m1603R, 2016JHEP...08..106M}.
The OTOC is motivated from the thermal expectation value of the commutator squared of operators $V,W$ separated in space and time with inverse temperature $\beta$
\begin{align}
    -\langle \left[V, W(t) \right]^2 \rangle_{\beta} = \langle V W(t) W(t) V \rangle_{\beta} + \langle W(t) VV W(t) \rangle_{\beta} -\langle V W(t) V W(t)\rangle_{\beta}-\langle W(t) V W(t) V \rangle_{\beta}.
\end{align}
The first two terms are time ordered and rather boring, and it is the latter out-of-time-ordered correlation functions that diagnose chaos. In particular, we will study
\begin{align}
    C_{\beta}(x,t) \equiv \frac{\langle V^{\dagger}W^{\dagger}(t) V W(t)\rangle_{\beta}}{\langle V^{\dagger}V\rangle_{\beta}\langle W^{\dagger} W \rangle_{\beta}}.
    \label{OTOC}
\end{align}
Qualitatively speaking, exponential decay in early time of the OTOC for generic operators provides a necessary condition for quantum chaos, where the decay rate is characterized by a quantum Lyapunov exponent. The correlator approaches a trivial value on the order of the ``scrambling time" for chaotic systems. The final value of the correlation function can be thought of as an overlap between states with different ordering of operator insertions. Thus, if (\ref{OTOC}) decays to zero, the states are very different and the butterfly effect has ensued, the late-time rate of decay indicating the strength of scrambling. In particular, this late-time behavior is what we mean by ``scrambling" in the context of OTOCs \cite{2014JHEP...03..067S,2015PhRvL.115m1603R}. This is distinct from another information-theoretic notion of scrambling that we will introduce in the next section, though under certain conditions, these may be shown to be related (see Fig.\ref{fig:cartoon}). Importantly, for certain systems, the correlator reaches an $O(1)$ constant, indicating that information is retained of the initial state.

Let's briefly review how to compute the OTOC in conformal field theory \cite{2015PhRvL.115m1603R}.  
Unlike Lorentzian field theory, in Euclidean field theory, there is only a single operator ordering within the correlation functions. 
% However, it is the Euclidean ones that are tractable.
We must therefore be careful with our analytic continuations to Lorentzian time where there are multiple ordering choices. 
We are concerned with the Euclidean thermal four-point function
\begin{align}
    \langle W(z_1,\bar{z}_1)W(z_2,\bar{z}_2)V(z_3,\bar{z}_3)V(z_4,\bar{z}_4) \rangle_{\beta} = \frac{1}{z_{12}^{2h_w}z_{34}^{2h_v}}\frac{1}{\bar{z}_{12}^{2\bar{h}_w}\bar{z}_{34}^{2\bar{h}_v}} f(z, \bar{z})
    \label{4point}
\end{align}
where $z = \frac{(z_1 - z_2)(z_3 - z_4)}{(z_1-z_3)(z_2-z_4)}$ is the cross ratio, 
and $f(z,\bar{z})$ depends on the full operator content of the theory.
The key point is that different analytic continuations to Lorentzian time correspond to different orderings of the operators in Lorentzian signature. We take 
\begin{align}
    z_1 &= e^{(2\pi/\beta)(t' + i \epsilon_1)}, \quad  z_2 = e^{(2\pi/\beta)(t' + i \epsilon_2)}, \quad  z_3 = e^{(2\pi/\beta)(x + i \epsilon_3)}, \quad  z_4 = e^{(2\pi/\beta)(x + i \epsilon_4)}, \nonumber \\
    \bar{z}_1 &= e^{-(2\pi/\beta)(t' + i \epsilon_1)}, \quad  \bar{z}_2 = e^{-(2\pi/\beta)(t' + i \epsilon_2)}, \quad  \bar{z}_3 = e^{(2\pi/\beta)(x - i \epsilon_3)}, \quad  \bar{z}_4 = e^{(2\pi/\beta)(x - i \epsilon_4)},   
\end{align}
such that the ordering of the operators in the correlation function from left to right correspond to smallest to largest $\epsilon_i$. At the end of the computation, the $\epsilon$'s should be sent to 0. When $t \gg x$,\
\begin{align}
    z \simeq -e^{(2\pi/\beta)(x-t)}\epsilon^*_{12}\epsilon_{34}, \quad \bar{z} \simeq -e^{-(2\pi/\beta)(x+t)}\epsilon^*_{12}\epsilon_{34},
    \label{OTOC_cross}
\end{align}
where we use the usual notation
\begin{align}
    \epsilon_{ij} \equiv i\left(e^{(2\pi/\beta)i \epsilon_i} - e^{(2\pi/\beta)i \epsilon_j}  \right).
\end{align}
Then, one just needs to  evaluate the Euclidean four-point function (\ref{4point}) with this parametrization.

\paragraph{Operator entanglement}
The entanglement measures of unitary time evolution operators capture the delocalization of information in quantum systems, and have been studied as an information-theoretic probe of chaotic dynamics in a variety of quantum systems \cite{2001PhRvA..63d0304Z,2007PhRvA..76c2316P,2009PhRvB..79r4416P,2016JHEP...02..004H,2017JPhA...50w4001D,2017PhRvB..95i4206Z,2018arXiv180300089J,2018arXiv181200013N,2019arXiv190607639K}.
% in SYK model, random unitary circuits, spin chains~\cite{2016JHEP...02..004H}, and in certain types of 2D CFTs~\cite{2018arXiv181200013N}.
Here we briefly review the construction of operator entanglement measures, in particular the bi- and tri-partite operator mutual information.

Under channel-state duality~\cite{CHOI1975285, JAMIOLKOWSKI1972275}, the time evolution operator acting on Hilbert space $\mathcal{H}$
\be
U_\epsilon (t) = e^{(-it - \epsilon) H} = \sum\limits_a e^{(-it - \epsilon) E_a} |a \rangle \langle a|
\ee
can be mapped to a state living in the Hilbert space space of operators on $\mathcal{H}$ which is isomorphic to the doubled Hilbert space $\mathcal{H}_1 \otimes \mathcal{H}_2$
\be
|U_\epsilon (t) \rangle = \mathcal{N} \sum\limits_a e^{(-it - \epsilon) E_a} |a\rangle_1 |a^*\rangle_2
\ee
where $\epsilon$ is a regulator, $\mathcal{N}$ is a normalization constant, and $|a \rangle$ is an eigenstate with energy $E_a$ of the Hamiltonian $H$. 
% The subscripts 1 and 2 refer to the doubled Hilbert spaces $\mathcal{H}_1$ and $\mathcal{H}_2$, each of which are copies of the original Hilbert space $\mathcal{H}$. 
We refer $\mathcal{H}_{1,2}$ as the input and output Hilbert spaces, respectively. 
Note that the regulator is only necessary when working with continuum field theories and the operator entanglement is useful for finite-dimensional systems as well.

\begin{figure}
     \centering
     \includegraphics[height = 4cm]{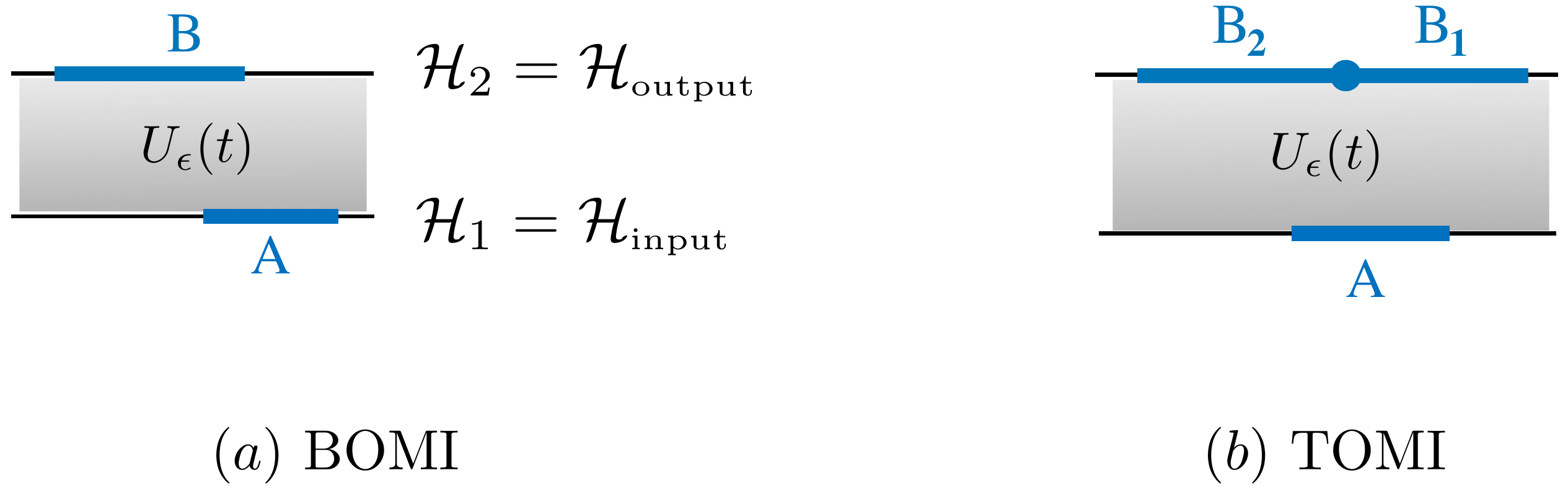}
     \caption{General setup of operator mutual information. $\mathcal{H}_{1,2}$ are the spatially 1-dimensional Hilbert spaces represented by the horizontal black lines. The subsystems are represented by the blue lines. The grey area in between the two Hilbert spaces indicates time evolution dictated by $U_\epsilon (t)$, with the arrow of time $t$ pointing from $\mathcal{H}_{\mbox{\tiny input}}$ to $\mathcal{H}_{\mbox{\tiny output}}$.
(a) Bi-partite case, (b) Tri-partite case. For TOMI, we generally take $B_1$ and $B_2$ to be a bipartition of the entire output Hilbert space.} 
\label{OMI_setup}
\end{figure}

Similar to constructing entanglement measures for the states living in the original Hilbert space $\mathcal{H}$, one can consider the bipartitioning of the total Hilbert space 
% $\mathcal{H}_{tot} = \mathcal{H}_1 \otimes \mathcal{H}_2$
into $\mathcal{H}_A$ and its complement $\mathcal{H}_{\bar A}$.
The $n^{th}$ R\'enyi operator entanglement entropy of subsystem $A$ is then defined by
\begin{align}
S^{(n)}_A = \frac{1}{1-n}\log [ \mbox{Tr}_A \rho_A^n],
\end{align}
where $\rho_A = \mbox{Tr}_{\bar A} | U_\epsilon (t) \rangle  \langle U_\epsilon (t) |$  is the reduced density matrix on $A$. With $S_A^{(n)}$, one can further introduce the $n$-th bi- and tri-partite operator mutual information (BOMI and TOMI), defined as linear combinations of entropies\footnote{The dependence of the R\'enyi index, $n$, for BOMI and TOMI is not a priori trivial. However, the kinematical factor (fixed by conformal symmetry) that describes the quasi-particle picture trivially depends on $n$ and violations of the quasi-particle picture at $n \neq 1 $ imply violations at $n = 1$. Furthermore, different R\'enyi entropies have been observed to act qualitatively similarly for operator entanglement in 2D CFTs \cite{2018arXiv181200013N,2019arXiv190607639K}. We then argue that our focus on $n=2$ in this paper captures the qualitative behavior of the other R\'enyi entropies (including the von Neumann entropy) in the theories of interest.}
\begin{align}
I^{(n)} (A,B) = S_A^{(n)} + S_B^{(n)} - S_{A \cup B}^{(n)},
\end{align}
and 
\begin{align}
I^{(n)}_3 (A,B_1, B_2) =  I^{(n)} (A,B_1) + I^{(n)} (A,B_2) - I^{(n)} (A,B_1 \cup B_2)
\label{TOMI_def}
\end{align}
respectively, for three sub-Hilbert spaces $\mathcal{H}_A, \mathcal{H}_{B_1}, \mathcal{H}_{B_2} \subset \mathcal{H}_1 \otimes \mathcal{H}_2$. Fig.~\ref{OMI_setup} shows the generic setup of BOMI and TOMI. Intuitively, BOMI characterizes the localization of information sent by the unitary channel from $A$ to $B$. BOMI of integrable theories, such as the free fermion, can be readily interpreted by the so-called quasi-particle picture where localized information carrying entities travel ballisticlly from $A$ to $B$~\cite{2005JSMTE..04..010C,2006PhRvL..96m6801C,2014PhRvL.112k1602N}. On the other hand, TOMI represents the delocalization of information scrambled by the unitary channel. In the context of operator entanglement, we are always referring to this delocalization when we say ``scrambling." In particular, a negative TOMI indicates that certain amount of information of $A$ is delocalized into the entire $B_1 \cup B_2$ after being processed by the unitary channel and cannot be extracted by simply measuring $B_1$ or $B_2$ separately~\cite{2016JHEP...02..004H, 2018arXiv181200013N}. By definition, the quasi-particle picture describes dynamical processes where no information is delocalized, so the TOMI is always trivial. However, in SYK and non-integrable spin chain models, the late-time saturation value of TOMI was found to approach the value of the Haar-random channel~\cite{2016JHEP...02..004H}. Furthermore, in holographic CFTs,
% for adjacent and semi-infinite $B_{1,2}$
the magnitude of late-time TOMI grows linearly with the size of $A$, in contrast with constant or logarithmic growth in certain non-holographic CFTs~\cite{2018arXiv181200013N}. In this work, we will present a more systematic way of understanding these scaling behaviors.

\subsection{Unification through the torus partition function}

Here, we remark on the interesting fact that the three quantities of interest may all be obtained directly from the torus partition function, the distinctions only coming from the analytic continuations of the modular parameters. We now make this explicit. This is a straightforward statement for the spectral form factor which may be defined in terms of the partition function
\begin{align}
    g(\beta, t) \equiv \left|\mathcal{Z}(\beta + i t)\right|^2.
\end{align}
The modular parameters have been continued as 
\begin{align}
    \tau \rightarrow \frac{i(\beta + i t)}{2\pi}, \quad \bar{\tau} \rightarrow -\frac{i(\beta + i t)}{2\pi}.
    \label{SFF_analytic cont}
\end{align}
% \textcolor{red}{[Make a remark on the $ST^2S$ transformation for integer time evolution on page 12 of \cite{2019JHEP...04..025B}.]}

Moving on to the OTOC, in general there is an arbitrary choice of operators to use.
It was argued in Ref.~\cite{2017PhRvD..96d6020C} that twist fields present a natural choice\footnote{We note that once we choose twist fields as our operators, the OTOC we are studying is no longer for the original CFT, but rather a replicated theory with couplings between the replicas. An implicit assumption of this paper is that studying OTOCs in the orbifold theory can tell us about the scrambling behavior of the seed theory. 
% We expect that if significant qualitative deviations occur, it will only be at large-N.
}. 
% The motivation for using twist fields as the operators in the OTOC is that the OTOC usually requires some arbitrary choice of primary operators to use. 
Choosing twist fields is natural because they are primary operators that are common to \textit{all} CFTs (after appropriate orbifolding). Moreover, we expect them to capture the main features of the seed theory because they naturally probe the entire spectrum of the seed theory through their relation to its torus partition function. While the OTOC of twist fields certainly is not always identical to the OTOC of the other primary fields, for the above arguments, we argue that they are a natural observable to study. 

In this case, the cross-ratios and hence modular parameters of the partition function must be analytically continued according to the prescription we have reviewed earlier. For the $\mathbb{Z}_2$ orbifold, the monodromy around the singular point at $z=1$ corresponds to a modular $S\bar{T}^2 S$ transformation \cite{2017PhRvD..96d6020C}
\begin{align}
    \tau \rightarrow \frac{\tau}{1+2\tau}, \quad \bar{\tau} \rightarrow \bar{\tau}.
\end{align}
where $\bar{T}$ is the inverse of the $T$ transformation i.e. $\tau \rightarrow \tau -1$. 
% {\color{blue} What is the meaning of
% $\bar{T}$ here?)} \textcolor{red}{[Clarified.]}
The anti-holomorphic modular parameter does not pick up a monodromy.
Because the twist-fields $\sigma_2$ are in the $\mathbb{Z}_2$ orbifold theory, there is no distinction from anti-twists $\bar \sigma_2$ and they both have conformal dimensions
\begin{align}
    h_{\sigma_2} = \bar{h}_{\sigma_2} = \frac{c}{16}.
\end{align}
% From the replica perspective, this correlation function is a partition function on a torus. 
It can be shown that the four-point function of twist fields on the complex plane may be equated with the torus partition function as follows
% For $n=2$, twist and anti-twist operators are the same and it was shown in 
\cite{2001CMaPh.219..399L} 
\begin{align}
    G_2(z, \bar{z}) \equiv \braket{\sigma_2(0)\bar \sigma_2(z, \bar{z}) \sigma_2(1) \bar \sigma_2(\infty)} = \left(2^8 z(1-z) \right)^{-c/24}\left(2^8 \bar{z}(1-\bar{z}) \right)^{-c/24}\mathcal{Z}(\tau, \bar{\tau})
    \label{eq: G2}
\end{align}
where $(\tau, \bar{\tau})$ are the modular parameters of the flat torus related to the cross-ratios as
% \begin{align}
%     x = \frac{\theta_2^4(\tau)}{\theta_3^4(\tau)}, \quad \bar{x} = \frac{\theta_2^4(\bar{\tau})}{\theta_3^4(\bar{\tau})}.
% \end{align}
% The inverse of these functions is
\begin{align}
    \tau = i \frac{K(1-z)}{K(z)}, \quad \bar{\tau } = -i \frac{K(1-\bar{z})}{K(\bar{z})}.
    \label{cross_mod}
\end{align}
Using \eqref{eq: G2}, the OTOC at late times is then 
\begin{align}
    C_{\beta} = 2^{-2c/3}\left| 1-z \right|^{-c/12} \left| z \right|^{c/6} \mathcal{Z}\left( \frac{\tau}{1+2\tau}, \bar{\tau} \right).
\end{align}
where the cross-ratios $z, \bar z$ are listed in (\ref{OTOC_cross}).
% , and the modular parameters are analytically continued to real time $t$ via
% \be
%  \tau = i \frac{K(1-z)}{K(z)}, \quad \bar{\tau } = -i \frac{K(1-\bar{z})}{K(\bar{z})}.
% \ee

We are left with the evaluation of operator entanglement. We specifically consider the second R\'enyi bipartite operator mutual information for intervals
\begin{align}
    A = [X_2 , X_1 ] , \quad B = [Y_2, Y_1].
\end{align}
with $A$ in the ``input" Hilbert space and $B$ in the ``output." The R\'enyi entropies of $A$ and $B$ separately are computed by two-point functions of twist-fields which are universal in CFT. Thus, the nontrivial contribution to the mutual information comes from the R\'enyi entropy of $A\cup B$ which may be computed by a four-point function of twist fields
% and anti-twist fields
% The OTOC and BOMI are both four-point functions. Choosing the operators to be twist-fields with $n=2$,
\begin{align}
    S^{(2)}_{A\cup B} &= -\log \langle \sigma_2(z_1, \bar{z}_1) {\sigma}_2(z_2, \bar{z}_2) \sigma_2(z_3, \bar{z}_3) {\sigma}_2(z_4, \bar{z}_4) \rangle \\ &= -\log \left(\left| z_{12} z_{34} \right|^{-c/2}\left| 1- x\right|^{-c/2}G_2(x ,\bar{x})\right),
\end{align}
where $x$ and $\bar{x}$ are the cross-ratios 
\begin{align}
    x = \frac{z_{12} z_{34}}{z_{13} z_{24}}, \quad \bar{x} = \frac{\bar{z}_{12} \bar{z}_{34}}{\bar{z}_{13} \bar{z}_{24}}.
\end{align}
(To avoid notation confusion with the analytically continued cross-ratios in OTOC shown in~\eqref{OTOC_cross}, here we opt for $x,\bar x$ as the cross-ratios.)

Up to a constant factor, the second R\'enyi operator mutual information is then \cite{2018arXiv181200013N}
\begin{align}
    I^{(2)}(A, B)(t) = \log\left[2^{-2c/3} \left|x \right|^{c/6}\left|1-x \right|^{-c/12}\mathcal{Z}(\tau, \bar{\tau}) \right].
    \label{eq: 2nd OMI}
\end{align}
 After analytic continuation to Lorentzian time, the cross-ratios are given by
\begin{align}
    x &= \frac{\sinh\left[\frac{\pi}{2 \epsilon}(X_1 - X_2) \right]\sinh\left[\frac{\pi}{2 \epsilon}(Y_1 - Y_2) \right]}{\cosh\left[\frac{\pi}{2 \epsilon}(X_1 - Y_2 -t)\right]\cosh\left[\frac{\pi}{2 \epsilon}(X_2 - Y_1 -t)\right]}, \nonumber \\
    \bar{x} &= \frac{\sinh\left[\frac{\pi}{2 \epsilon}(X_1 - X_2) \right]\sinh\left[\frac{\pi}{2 \epsilon}(Y_1 - Y_2) \right]}{\cosh\left[\frac{\pi}{2 \epsilon}(X_1 - Y_2 +t)\right]\cosh\left[\frac{\pi}{2 \epsilon}(X_2 - Y_1 + t)\right]}
    \label{OMI_cross}
\end{align}
The TOMI is then computed as the linear combination \eqref{TOMI_def}.

\subsection{Summary of results}

We summarize our results of studying the torus partition function in Table \ref{results_table}, pointing to relevant references when appropriate.
\begin{table}
\begin{center}
\scriptsize
% \textit{Italicized font denotes original work from this project.} \textcolor{red}{[Needs to be updated]}
  \begin{tabular}{||p{.125\textwidth}|p{.275\textwidth}|p{.275\textwidth}|p{.275\textwidth}||}
  
    \hline
    % \rowcolor{LightCyan}
     & \textbf{Spectral form factor} & \textbf{OTOC} & \textbf{Operator entanglement} \\ \hline
    \textbf{Free fermion}  & Periodic with $t_{rec} = 2\pi$ (Sec.~\ref{sec_free_sff}). & Saturation to a constant $C_{\beta} \rightarrow \frac{ \left(e^{i\pi/12}+e^{-i\pi/6}\right)}{2}$ (Sec.~\ref{sec_free_OTOC}).& Quasi-particle picture $I^{(2)}_3 = 0$ \cite{2018arXiv181200013N} (Sec.~\ref{sec_free_TOMI}).\\ \hline
    \textbf{RCFT} & Emergence of dip, (non-linear) ramp, and plateaux in large $m$ limit of unitary minimal models $\mathcal{M}(m,m+1)$ \cite{2019JHEP...04..025B} and large $k$ limit of $\mathfrak{su}(2)_k$ WZW models (Sec.~\ref{MM_SFF} \& \ref{su2_SFF}). & Saturation to constant $(S\bar{T}^2 S)_{00}$. This constant decays to zero for large $m$ ($\mathcal{M}(m,m+1)$) and $k$ ($\mathfrak{su}(2)_k$) (Sec.~\ref{MM_OTOC} \& \ref{su2_OTOC}). This may be compared to generic results for RCFTs \cite{2016JHEP...08..129G, 2016PTEP.2016k3B06C} . &Approximate quasi-particle picture with constant late-time TOMI related to modular data $I^{(2)}_3 = \log \frac{S_{00}^2}{\sum_h S_{0h}^2}$ (Sec.~\ref{sec_RCFT}).  \\ \hline
    
    \textbf{Irrational CFT}  & Infinite recurrence time. Suspected to mimic random matrix theory with dip, ramp, and plateau (Sec.~\ref{sec_ICFT}).
    & Exponential decay as $ e^{-\frac{\pi (c-1)t}{12\beta} }t^{-\frac{3}{2}}$ \cite{2019arXiv190502191K} (Sec.~\ref{sec_ICFT}).& Scrambling restricted by the number of conserved currents $0 \leq  I^{(2)}_{AB_1}, I^{(2)}_{AB_2} \leq \frac{\pi l (2c_{currents}+c)}{24 \epsilon }+ \frac{c_{currents}-c}{3}\log 2+ \log S_{00}$ (Sec.~\ref{sec_ICFT}). \\ \hline
    
    \textbf{Compactified Boson ($\eta \notin \mathbb{Q}$)}  &  Dip and plateau with no clear ramp. Infinite recurrence time (Sec.~\ref{CB_SFF}).
    &  Polynomial decay to zero \cite{2017PhRvD..96d6020C} $C_{\beta} = -\frac{\pi}{2 \log\left( -\frac{\epsilon_{12}*\epsilon_{34}}{16}e^{-\frac{2\pi(t-x)}{\beta}}\right)}$ (Sec.~\ref{CB_OTOC}).& Approximate quasi-particle picture, though nontrivial TOMI that scales logarithmically with system size \cite{2018arXiv181200013N} $I^{(2)}_3 \sim -\log l/\epsilon $ (Sec.~\ref{CB_TOMI}).\\ \hline
    
    \textbf{Holographic CFTs}  & Dip, ramp, and plateau nearly recovered by summing over gravitational saddles beyond thermal $AdS$ and BTZ black hole \cite{2017JHEP...08..075D}. Reviewed in (Sec.~\ref{holo_SFF}). &
    Exponential decay as $\propto e^{-2\pi c t/12\beta}$ contrary to the story for heavy-light OTOC  $\propto e^{-2\pi \Delta_{\mathcal{O}}t/\beta}$. Thermal $AdS$ saddle dominates (Sec.~\ref{holo_OTOC}). 
    % Studied extensively: Lyapunov behavior \cite{2015PhRvL.115m1603R,2014JHEP...03..067S,2016JHEP...08..106M}.
    & Strong violation of quasi-particle picture with TOMI scaling extensively but \textit{not} saturating the theoretical bound $I^{(2)}_3 \rightarrow -\frac{\pi c l}{4\epsilon}$. Both BTZ black hole and Thermal $AdS$ saddles contribute (Sec.~\ref{holo_TOMI}).
    \\ \hline

\end{tabular}
\end{center}
\caption{Summary of results and related works.}
% {\color{red}Caption}}
\label{results_table}
\end{table}

The organization of the paper is as follows. In Section \ref{sec_free}, we warm up by considering free theories, setting a baseline for simple integrable behavior. In Section \ref{sec_RCFT}, we progress to generic rational CFTs. Though rational, we find an
% {\color{blue} an (?)}
emergent irrational structure as we complicate the theories by systematically introducing more highest weight representations. As concrete examples, we study the large $m$ limit of the unitary minimal models $\mathcal{M}(m,m+1)$ and the large $k$ limit of $\mathfrak{su}(2)_k$ Wess-Zumino-Witten models. In Section \ref{sec_ICFT}, we take the leap to irrational CFTs characterized by central charges larger than 1 and no extended symmetry algebra beyond Virasoro. After explaining generic results for irrational CFT, we study the compactified boson at irrational radius; despite having central charge $c=1$, this theory is not quite rational due to the infinite number of highest weight representations and thus displays some interesting features of the breakdown of rationality. In Section \ref{sec_HCFT}, we study conformal field theories with semiclassical holographic duals, which are known to be ``maximally chaotic." In particular, the OTOC of twist-fields corresponds to four heavy operators, probing a different regime of scrambling than the heavy-light limit usually studied. Consequently, we find a different exponential scaling. Finally, in Section \ref{sec_discussion}, we discuss future directions in unifying the chaos web.

%%%%%%%%%%%%%%%%%%%%%%%%%%%%%%%%%
\section{Free theories}
\label{sec_free}
We warm up by considering free theories where the absence of interactions implies that quantum information remains localized, as do Heisenberg operators. This will set a baseline for the diagnostics that we are studying. To be explicit, we study the free massless fermion with torus partition function
\begin{align}
    \mathcal{Z}\left( \tau ,\bar{\tau} \right)= \frac{1}{2}\left(\left| \frac{\theta_2(\tau)}{\eta(\tau)}\right|+\left| \frac{\theta_3(\tau)}{\eta(\tau)}\right|+\left| \frac{\theta_4(\tau)}{\eta(\tau)}\right|\right),
\label{FFPartitionFunction}
\end{align}
where $\theta_i(\tau) \equiv \theta_i(0 | \tau)$, $\theta_i(z| \tau)$ are theta functions\footnote{We use the conventions from Ref.~\cite{DiFrancesco:1997nk} which we note are different from those in Mathematica and Python.} and $\eta$ is the Dedekind eta function. 

\subsection{SFF}
\label{sec_free_sff}
\begin{figure}
    \centering
    % \begin{subfigure}
    % \centering
        \includegraphics[width=0.48\linewidth]{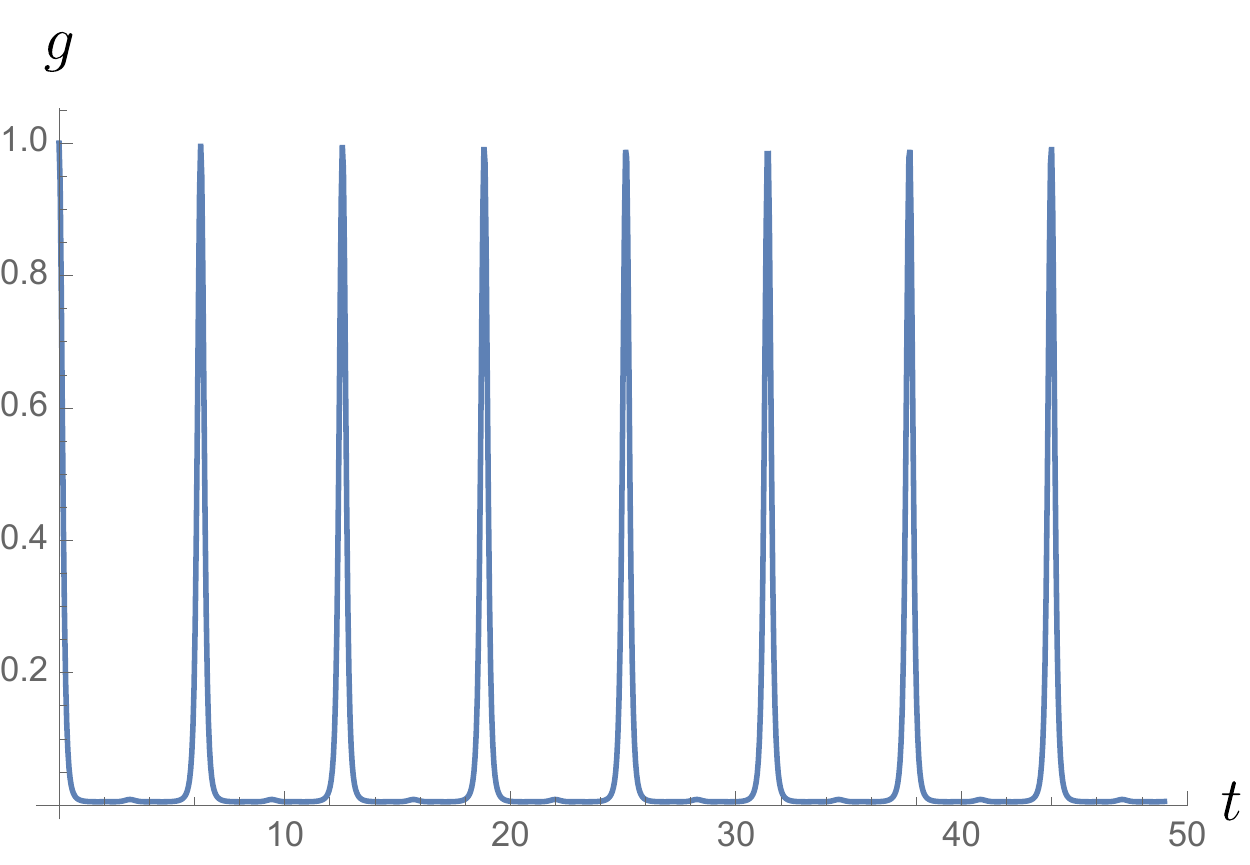}
        \caption{Spectral form factor for a single free fermion with $\beta = .5$. The function is clearly periodic and does not demonstrate anything close to random matrix statistics. The recurrence time is  $2\pi$. }
    \label{fig: FF_SFF}
\end{figure}

From the expression~\eqref{FFPartitionFunction} and the analytic continuation~\eqref{SFF_analytic cont}, it is straightforward to see that the SFF of free fermion has a recurrence time
\be
t_{rec} = 2\pi.
\ee
This is independent of the inverse temperature $\beta$. We show this in Fig.~\ref{fig: FF_SFF}. 
As pointed out in Ref.~\cite{2019JHEP...04..025B}, one should restrict analysis to the time regime $t < t_{rec}$ in order to properly discuss if there is any random matrix theory-like behavior. Due to the regular sinusoidal feature and small recurrence time, we obviously detect no chaotic spectral statistics.

\subsection{OTOC}
\label{sec_free_OTOC}
The theta and Dedekind eta functions have simple modular properties such that the partition function transforms as
\begin{align}
    \mathcal{Z}\left( \tau ,\bar{\tau} \right) \rightarrow \frac{1}{2}\left(e^{i\pi/12}\left| \frac{\theta_2(\tau)}{\eta(\tau)}\right|+e^{i\pi/12}\left| \frac{\theta_3(\tau)}{\eta(\tau)}\right|+e^{-i\pi/6}\left| \frac{\theta_4(\tau)}{\eta(\tau)}\right|\right)
\end{align}
under the $S\bar{T}^2 S$ transformation. In the late-time limit, we can take the leading order terms in the series
\begin{align}
    \eta(q) \simeq q^{1/24}, \quad \theta_2(q) \simeq 2 q^{1/8}, \quad \theta_3(q) \simeq 1 + 2 q^{1/2}, \quad \theta_4(q) \simeq 1 -2q^{1/2}
\end{align}
such that 
\begin{align}
    \mathcal{Z}\left( \frac{\tau}{1+2\tau} ,\bar{\tau} \right) \simeq \frac{\left(e^{i\pi/12}+e^{-i\pi/6}\right) }{2}\left|q \right|^{-1/24} \simeq \frac{\left(e^{i\pi/12}+e^{-i\pi/6}\right) }{2}\left|\frac{z}{16} \right|^{-1/12}.
\end{align}
Because $c=1/2$ for the free fermion, we find
\begin{align}
    C_{\beta} \rightarrow \frac{ \left(e^{i\pi/12}+e^{-i\pi/6}\right)}{2}
\end{align}
which is a constant with absolute value very close to the early time value of unity, so little information has been scrambled in the sense that was put forward in the introduction. We attribute this minimal scrambling as an artifact of the coupling of the the two copies of the theory induced by the orbifolding. We stress that this notion of information scrambling is distinct from that of the late-time value of $I_3$.

\subsection{TOMI}
\label{sec_free_TOMI}

\begin{figure}
    \centering
    \includegraphics[width = \textwidth]{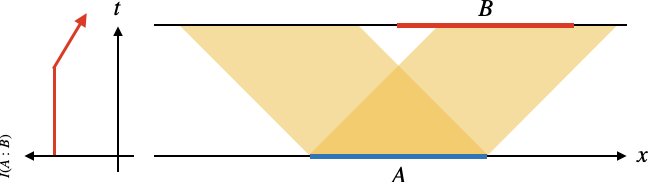}
    \caption{A cartoon of the quasi-particle picture for operator entanglement. At $t = 0$, the quasi-particles (yellow) are emitted from region $A$ and move at the speed of light. The quasi-particle picture for operator entanglement dictates that the operator mutual information between input region $A$ and output region $B$ is the number of quasi-particles located within region $B$ that originated from $A$ (the overlap section between region B and the yellow area). On the left, we display a sketch of the operator mutual information between $A$ and $B$ for the partially overlapping configuration: after a certain time $I^{(2)}(A,B)$ starts to decrease due to the quasi-particles leaving region $B$.}
    \label{quasi_cartoon}
\end{figure}

As a free (trivially integrable) theory, the free fermion is known to be well-described by the quasi-particle picture \cite{2006PhRvL..96m6801C,2014PhRvL.112k1602N}.
% {\color{red} [the quasi particle picture in Refs [28] - [32] of 1501.00568 is not exactly the same?]}
Because the quasi-particle picture describes infinitely-living localized objects carrying the entanglement, no information is ever delocalized. The quasi-particle picture for operator entanglement is modified from that of a global quench but is morally the same (see Fig.~\ref{quasi_cartoon}). This was confirmed for operator entanglement in CFT for both mutual information and logarithmic negativity in Refs.~\cite{2018arXiv181200013N,2019arXiv190607639K}. It was found using twist operators that the second R\'enyi bipartite operator entanglement is 
\begin{align}
\label{fOMI}
  &I^{(2)}(A,B)(t) =  \frac{c}{8}\log \left(\frac{ \cosh\frac{\pi(Y_2-X_1+t)}{2\epsilon} \cosh\frac{\pi(Y_1-X_2+t)}{2\epsilon} \cosh\frac{\pi(Y_2-X_1-t)}{2\epsilon} \cosh\frac{\pi(Y_1-X_2-t)}{2\epsilon} }{ \cosh\frac{\pi(Y_2-X_2+t)}{2\epsilon} \cosh\frac{\pi(Y_1-X_1+t)}{2\epsilon} \cosh\frac{\pi(Y_2-X_2-t)}{2\epsilon} \cosh\frac{\pi(Y_1-X_1-t)}{2\epsilon}} \right).
\end{align}

Equivalently, we could have employed the expression involving the torus partition function (\ref{eq: 2nd OMI}). This leads to a trivial tripartite operator mutual information
\begin{align}
    I_3^{(2)}(t) = 0,\quad \forall t.
\end{align}
Thus, the unitary evolution operator for free fermions does not delocalize any information, as predicted by the quasi-particle picture.

% \begin{figure}
%     \centering
%     \includegraphics[height = 5cm]{OMI_free_bos.pdf}
%     \caption{Bipartite operator mutual information for intervals separated by 50 with $|A| = 100, |B| \in \{100,180,260 \}$, and $\epsilon = 1$. This quasi-particle picture is evident but there the value becomes negative at early and late times which should not be happening. \textcolor{red}{[Negative value seems to go away in the scaling limit but it is still strange.]}}
%     \label{bos_OMI}
% \end{figure}

\section{RCFT}
\label{sec_RCFT}

We now include interactions by considering generic rational CFTs; these are conformal field theories with a finite number of representations of the Virasoro algebra, hence the partition function is a finite sum over characters\footnote{We have chosen the diagonal modular invariant for simplicity, but non-diagonal theories should be straightforwardly generalizable.}
\begin{align}
    \mathcal{Z} = \sum_{h} \chi_{h}(q)\bar{\chi}_h(\bar{q}).
    \label{Z_RCFT}
\end{align}
Before investigating specific families of RCFTs, we first make some universal analysis directly from the structure of the partition function.

\paragraph{OTOC} It has been generically shown that the late-time behavior of OTOCs in RCFTs is a constant value equal to the $00$ component of the monodromy matrix \cite{2016JHEP...08..129G,2016PTEP.2016k3B06C}
\begin{align}
    C_{\beta} (x,t) \rightarrow \textbf{M}_{00} = \frac{1}{d_i d_j}\frac{S_{ij}^*}{S_{00}},
\end{align}
where $d_{i,j}$ are the quantum dimensions of the operators in the OTOC and $S_{ij}$ is the matrix element of the modular matrix. Thus, the OTOC of twist fields in the orbifold theory is
\begin{align}
    C_{\beta}(x,t) \rightarrow \frac{1}{d_{\sigma_2}^2}\frac{S_{\sigma_2 \sigma_2}^*}{S_{00}}.
    \label{C_orbi_eq}
\end{align}
As we are unable to evaluate these modular matrix elements for orbifold CFTs, we directly compute the late-time OTOC from the torus partition function; this may be viewed as a way to extract modular data for the orbifold theory.

% We can derive a similar result directly from the partition function (\ref{Z_RCFT}) . 
For the OTOC we are concerned with, the $z, \bar{z} \rightarrow 0$ limit corresponds to late times. Thus, only the vacuum character will contribute and the modular transformed partition function reduces to
\begin{align}
    \mathcal{Z} \rightarrow (S\bar{T}^2 S)_{00} q^{-c/24}\bar{q}^{-c/24} = (S\bar{T}^2 S)_{00} z^{-c/12}\bar{z}^{-c/12} 2^{2c/3}
    \label{xxbar00}
\end{align}
Thus, the OTOC is\footnote{It is interesting to note that this is related to the square root of the spectral form factor at time $t=2\pi$ which can be seen by applying discrete time evolution to the partition function \cite{2019JHEP...04..025B}.}
% \textcolor{red}{[check factors with (3.13) of Takayanagi]}
\begin{align}
    C_{\beta} \rightarrow (S\bar{T}^2 S)_{00} .
    \label{rcft_otoc_eq}
\end{align}
Note that this modular data is for the seed theory, not the orbifold theory. Thus, it is more readily computable and we will work this out explicitly for specific models.
\eqref{C_orbi_eq} and \eqref{rcft_otoc_eq} imply a consistency condition in modular data for orbifold CFTs and their seed theories 
% \textcolor{red}{[check]}
\begin{align}
    \frac{1}{d_{\sigma_2}^2}\frac{S_{\sigma_2 \sigma_2}^*}{S_{00}}\Bigg|_{orbifold} =  (S\bar{T}^2 S)_{00}\Big|_{seed}.
\end{align}

\paragraph{TOMI} For the operator entanglement, we must consider different limits of the modular parameters depending on the configurations of the intervals. 
We will consider the case where $B_1$ and $B_2$ are adjacent and spatially semi-infinite.
For definiteness, let us take $A = (0,l),$ $B_1 = (-\infty, 0),$ and $B_2 = (0,\infty)$, though the late-time behavior will end not depending on these positions, only on the length of $A$. The cross-ratios for the different subregions are
\begin{align}
    x_{AB_1}&\rightarrow  e^{\pi (t-l)/2\epsilon} \frac{\sinh \frac{\pi l }{2\epsilon}}{\cosh \frac{\pi t}{2 \epsilon}}, \quad \bar{x}_{AB_1} \rightarrow e^{-\pi (l+t)/2\epsilon} \frac{\sinh \frac{\pi l }{2\epsilon}}{\cosh \frac{\pi t}{2 \epsilon}}
\nonumber
    \\
    x_{AB_2}&\rightarrow  e^{-\pi t/2\epsilon} \frac{\sinh \frac{\pi l }{2\epsilon}}{\cosh \frac{\pi (l-t)}{2 \epsilon}}, \quad \bar{x}_{AB_2} \rightarrow e^{\pi t/2\epsilon} \frac{\sinh \frac{\pi l }{2\epsilon}}{\cosh \frac{\pi (l+t)}{2 \epsilon}}
\nonumber
    \\  
    x_{AB}&\rightarrow 1 - e^{-\pi l/\epsilon}, \quad \bar{x}_{AB} \rightarrow 1 - e^{-\pi l/\epsilon}
    \label{crosses_OMI}
\end{align}

In the limit that $x, 1-\bar{x} \rightarrow 0$, needed for $I_{AB_2}^{(2)}$, the sum will be dominated by the vacuum chiral character, but we must still sum over any anti-chiral currents \cite{2015JHEP...09..110A}
\begin{align}
    \mathcal{Z} = q^{-c/24}\sum_{\bar{h} \in currents}\bar \chi_{\bar{h}}(\bar{q}).
\end{align}
By defining a Cardy-like asymptotic density of currents a
\begin{align}
    \rho_{currents}(\bar{L}_0) \sim \exp \left( 2\pi \sqrt{\frac{c_{currents}\bar{L}_0}{6}}\right), \quad  \bar{L}_0 \rightarrow \infty
\end{align}
one can find \cite{2015JHEP...09..110A}
\begin{align}
    \mathcal{Z} \rightarrow 2^{(c+c_{currents})/3}x^{-c/12}(1-\bar{x})^{-c_{currents}/12} S_{00}.
\end{align}
Analogously, in the limit that $\bar{x}, 1-x\rightarrow 0$, needed for $I_{AB_1}^{(2)}$,
\begin{align}
    \mathcal{Z} \rightarrow 2^{(c+c_{currents})/3}(1-x)^{-c_{currents}/12}\bar{x}^{-c/12}S_{00}.
\end{align}
The final limit that we need is $x,\bar{x}\rightarrow 1$ which may be obtained by S-transformations on both of the characters such that %partition function such that
\begin{align}
    \mathcal{Z} &\simeq \sum_{h} S_{h0}^2 \chi_{0}(-1/\tau)\bar{\chi}_0(-1/\bar{\tau})
\\
&\rightarrow 2^{2c/3} (1-x)^{-c/12}(1-\bar{x})^{-c/12} \sum_{h} S_{h0}^2
\end{align}
We may then obtain general expressions for the saturation values of the BOMI and TOMI using \eqref{eq: 2nd OMI}. The BOMIs are
% \begin{align}
%      I^{(2)}(t) &= \log \left[\left| x\right|^{c/6}\left|2^8 (1-x) \right|^{-c/12} \mathcal{Z} (x,\bar{x})\right] 
% \end{align}
% We find
\begin{align}
    I^{(2)}_{AB_1} &=  \log \left[S_{00} x^{c/12}(1-\bar{x}) ^{-c/24} 2^{(c_{currents}-c)/3}(1-x)^{-(2c_{currents}+c)/24}\right]  \nonumber
    \\
    &\rightarrow \frac{\pi l (2c_{currents}+c)}{24 \epsilon } + \frac{c_{currents}-c}{3}\log 2+ \log S_{00},
\\
    I^{(2)}_{AB_2} &\rightarrow \frac{\pi l (2c_{currents}+c)}{24 \epsilon }+ \frac{c_{currents}-c}{3}\log 2+ \log S_{00},
\\
    I^{(2)}_{AB} &= \log \left[\left| x_{AB}\right|^{c/6}\left|1-x_{AB} \right|^{-c/4} \sum_{h} S_{h0}^2\right]= \frac{c\pi l}{4\epsilon} + \log \sum_{h} S_{h0}^2.
\end{align}
Thus, the TOMI is
\begin{align}
    I^{(2)}_3 \rightarrow \frac{\pi l (c_{current}-c)}{6\epsilon }+ \frac{2(c_{currents}-c)}{3}\log 2 + \log \frac{S_{00}^2}{\sum_h S_{h0}^2}.
    \label{I3_RCFT}
\end{align}
For all rational CFTs, $c_{current} = c$, so we generically recover a constant
\begin{align}
    I^{(2)}_3 = \log \frac{S_{00}^2}{\sum_h S_{h0}^2}.
    \label{RCFT_I3}
\end{align}
% \textcolor{red}{[We can evaluate this explicitly for minimal models and WZW.]}
This recovers the perfect quasi-particle picture when $S_{00}$ dominates the sum in the denominator, as was demonstrated for the free fermion. More complicated RCFT's may receive subleading contributions that lead us to a nontrivial constant. Any nontrivial constant of TOMI is a minor violation of the quasi-particle picture because the quasi-particle picture only describes the bipartite entanglement.
Furthermore, the number of currents is always bounded by the central charge so (\ref{I3_RCFT}) is negative semi-definite. Interestingly, it shows that the linear scaling of the tripartite mutual information may be somewhat generic, appearing in non-holographic theories without maximal chaos. %like holographic theories. 
We will return to this discussion in Sec.~\ref{sec_ICFT}.

\subsection{Minimal Models}

The so-called ``minimal models" represent the canonical family of RCFTs \cite{Belavin:1984vu}. This family of theories is parametrized by two co-prime integers $p,p'$. However, in order to satisfy unitarity, we restrict to $p' = p-1 \equiv m$ with $m\geq3$. The central charge lies between $1/2$ (for the Ising model) and $1$ (for the Runkel-Watts theory \cite{2001JHEP...09..006R})
\be
 c = 1- \frac{6}{m(m+1)},
\ee
and the conformal dimensions of the primary fields are 
\begin{align}
    h_{r,s} = \frac{((m+1)r-ms)^2-1}{4m(m+1)}, \quad 1 \leq r < m-1, \quad 
    1\leq s < m.
\end{align}
% Here we investigate the operator mutual information of minimal models. In particular, we will work with the second R\'enyi operator mutual information because this can be computed by torus partition functions. 
% We now review the relevant equations for computing partition functions of the diagonal minimal models following chapter 10 of~\cite{DiFrancesco:1997nk}. 
% The central charge of minimal models is parameterized by two co-prime integers $p,p'$:
% \be
%  c = 1- 6\frac{(p - p')^2}{p p'}.
% \ee
The torus partition function for diagonal minimal models is
\begin{align}
    \mathcal{Z}(\tau, \bar{\tau}) = \sum_{(r,s) \in E_{m+1,m}} \chi_{r,s}(\tau) \bar{\chi}_{r,s}(\bar{\tau})
    \label{eq: minimal models torus Z}
\end{align}
where $E_{m+1,m}$ is the set of pairs $(r,s)$ such that
\begin{align}
     ms< (m+1)r.
\end{align}
The characters of the minimal models are known
\begin{align}
    \chi_{r,s}(\tau) = K_{\lambda_{r,s}}-K_{\lambda_{r,-s}}, \quad \lambda_{r,s} = (m+1)r - ms
    \label{eq:mmcharacters1}
\end{align}
where 
\begin{align}
    K_{\lambda}(\tau) = \frac{1}{\eta(\tau)}\sum_{n \in \mathbb{Z}}q^{(Nn+\lambda)^2/2N}, \quad     q = e^{2 \pi i \tau}, \quad N = 2m(m+1).
    \label{eq:mmcharacters2}
\end{align}

\subsubsection{SFF}
\label{MM_SFF}

The spectral form factor of unitary diagonal minimal models was extensively studied in Ref.~\cite{2019JHEP...04..025B}. Here, we briefly reproduce their results for a representative set of parameters, but instead of ordinary time averaging or averaging over a window of $m$, we perform progressive time averaging introduced in Sec.~\ref{sec: intro} that effectively treats the rapid oscillations in the late time. 

\begin{figure}
    \centering
    % \begin{subfigure}
    % \centering
        % \includegraphics[width=0.48\linewidth]{MM_SFF_NonLogLog_m5_090419.pdf}
        %  \includegraphics[width=0.48\linewidth]{MM_SFF_P&OTA_m5_090419.pdf}
        % \caption{}
    % \end{subfigure}
    % \begin{subfigure}
    % \centering
        % \includegraphics[width=0.48\linewidth]{MM_SFF_NonLogLog_m50_090419.pdf}
        %  \includegraphics[width=0.48\linewidth]{MM_SFF_P&OTA_090319.pdf}
        \includegraphics[width=0.48\linewidth]{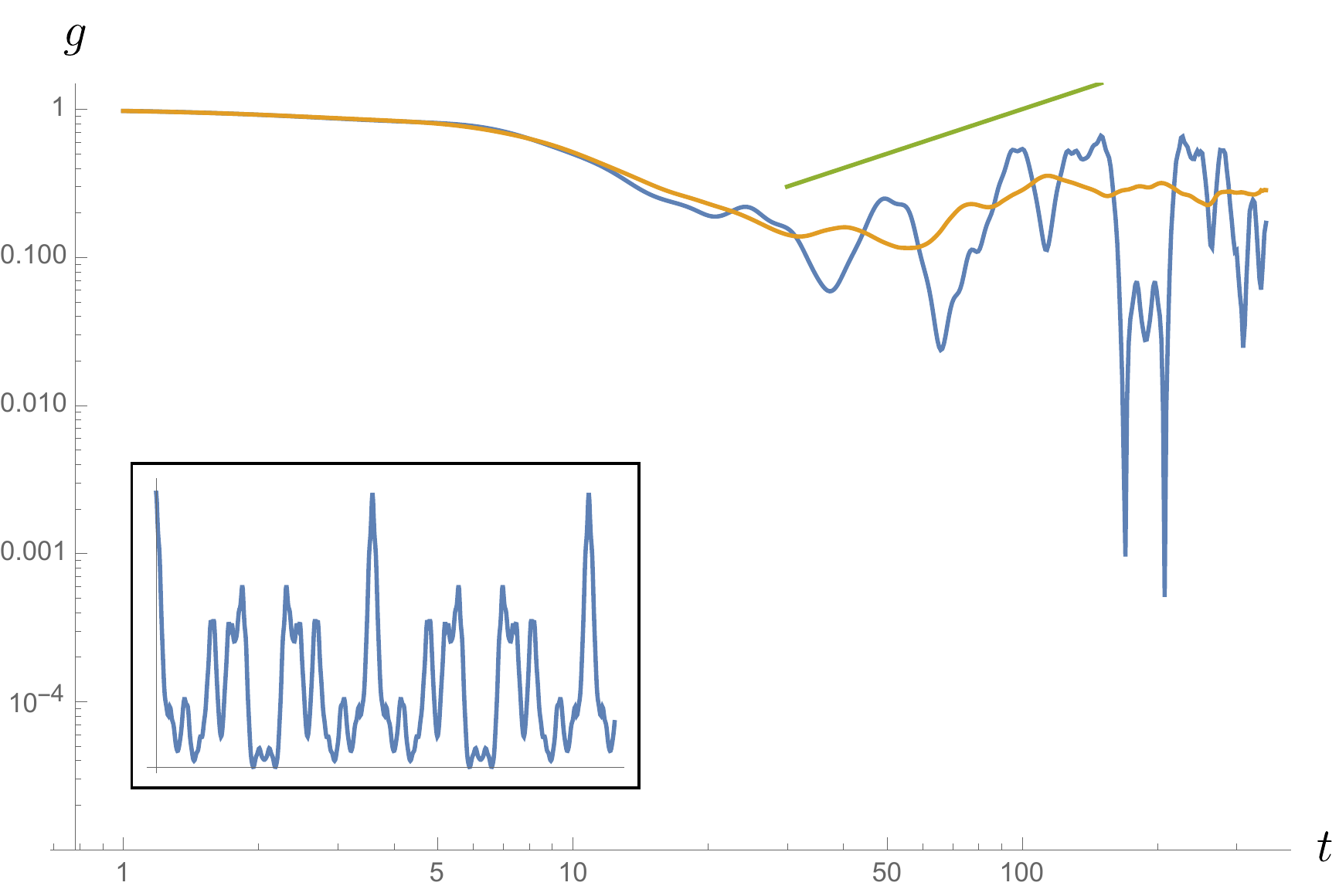}
        \includegraphics[width=0.48\linewidth]{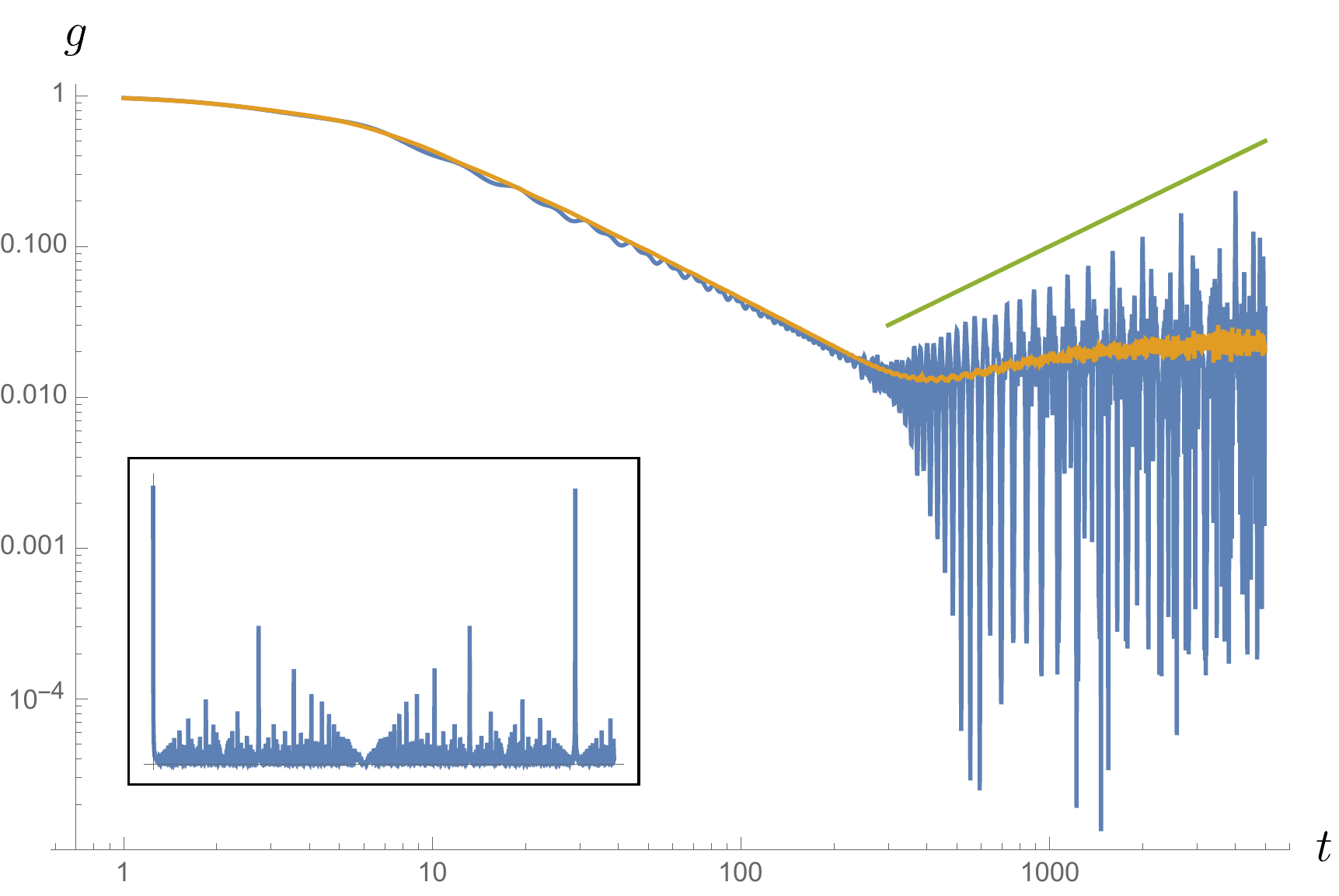}
        %  \caption{}
    % \end{subfigure}
        \caption{The spectral form factor (blue) and its progressive time-averaged version (yellow) for the unitary minimal models with $m = 5$ (left) and $m=50$ (right). We see the emergence of the dip, ramp, and plateau as we increase $m$.
        % Left: recurrence in non-log-log plot. Right: log-log plot for $t < t_{rec}$. $m = 50$. Left: recurrence in non-log-log plot. Right: log-log plot for $t < t_{rec}$. 
        While the peaks in the non-averaged function grow linearly, the time-averaged function is sublinear. This can be seen by the green lines which are linear in $t$. The insets show the functions on linearly scaled axes.
        % {\color{blue} (The green lines in these figures should be explained. 
        % Also, we perhaps need to optimize where to put these.)}\textcolor{red}{[Explained.]}
        }
    \label{fig: MM_SFF}
\end{figure}

The recurrence time of minimal models is $t_{rec} = 4 \pi m(m+1)$~\cite{2019JHEP...04..025B}, so for the dip-ramp-plateau feature to emerge before $t_{rec}$, $m$ must be taken to be large. Fig.~\ref{fig: MM_SFF} shows that $p = 50$ provides enough room for the SFF to begin to reveal a dip-ramp-plateau structure. In Ref.~\cite{2019JHEP...04..025B}, it was shown that the peaks of the spectral form factor grow linearly in time. However, these peaks become rarer and rarer as $t$ increases, so the averaged form factor has a sublinear ramp. Therefore, we believe that the ramp cannot be associated to the spectral rigidity of random matrix theory. With the assumption that random matrix theory provides the most rigid spectral statistics, we conclude that the nonlinear ``ramps" that we observe imply less rigid spectra.

\subsubsection{OTOC}
\label{MM_OTOC}

% For RCFTs, we follow the approach from Ref.~\cite{2017PhRvD..96d6020C}. RCFTs have a finite number of representations so their partition function may be written as a finite sum of characters
% \begin{align}
%     \mathcal{Z}(\tau, \bar{\tau}) = \sum_{\lambda = 0}^{N-1} \chi_{\lambda}(\tau)\bar{\chi}_{\lambda}(\bar{\tau})
% \end{align}
% where $N$ is the number of primaries and we have taken the diagonal modular invariant. We apply the $S\bar{T}^2S$ modular transformation
% \begin{align}
%     \mathcal{Z}\left(\frac{\tau}{1+2\tau }, \bar{\tau}\right) =\sum_{\nu = 0}^{N-1} \sum_{\lambda = 0}^{N-1} (S\bar{T}^2S)_{ \lambda\nu}\chi_{\nu}(\tau)\bar{\chi}_{\lambda}(\bar{\tau})
% \end{align}

For the minimal models, the explicit form of the modular matrices are known 
% \cite{DiFrancesco:1997nk}
\begin{align}
\label{MM_T}
    T_{rs;\rho\sigma} &= \delta_{r,\rho}\delta_{s,\sigma}e^{2\pi i(h_{r,s}-c/24)},
    \\
    S_{rs;\rho\sigma} &= 2\sqrt{\frac{2}{m(m+1)}}(-1)^{1 + s\rho+ r\sigma} \sin\left(\pi \frac{m+1}{m}r \rho \right)\sin\left(\pi \frac{m}{m+1}s\sigma \right),
    \label{MM_S}
\end{align}
so the full transformation needed for evaluating the OTOC is
\begin{align}
    (S\bar{T}^2S)_{11;11} = {\frac{8}{m(m+1)}} \sum_{(r,s) \in E_{m+1,m}} \sin^2\left(\pi \frac{m+1}{m}r \right)\sin^2\left(\pi \frac{m}{m+1}s\right) e^{-4i\pi(h_{r,s}-c/24)}.
\end{align}
Our argument that led to \eqref{rcft_otoc_eq} was rather quick, so we explicitly check this result for minimal models.
%  \bea
%  \mathcal{Z}_{p/p'}(\frac{\tau}{1+2\tau}, \bar \tau)  &=& \sum\limits_{\lambda= 0}^{N-1}  \sum\limits_{\nu= 0}^{N-1} (S \bar T^2 S)_{\lambda \nu} K_{\nu}(\tau) K_{\omega_0 \lambda} (\bar \tau)   \nonumber\\
%  &=& \sum\limits_{\lambda= 0}^{N-1}  \sum\limits_{\nu= 0}^{N-1} (S \bar T^2 S)_{\lambda \nu}    q^{-\frac{1}{24}}    \bar q^{-\frac{1}{24}}   q^{\frac{\lambda^2}{2N}}  \bar q^{\frac{\nu^2}{2N}} \nonumber\\
%  &=& (S \bar T^2 S)_{00}  q^{-\frac{1}{24}}    \bar q^{-\frac{1}{24}}
%  \eea
%  where in the last step we dropped all the terms except $\lambda = \nu = 0$ because again, $q \to 0$. This is Eq. (3.11) in \cite{2017PhRvD..96d6020C}.
% %\end{comment}
% %%%%%%%%%%%%%%%%%%%%%%%%

\begin{figure}
    \centering
    \includegraphics[height = 5cm]{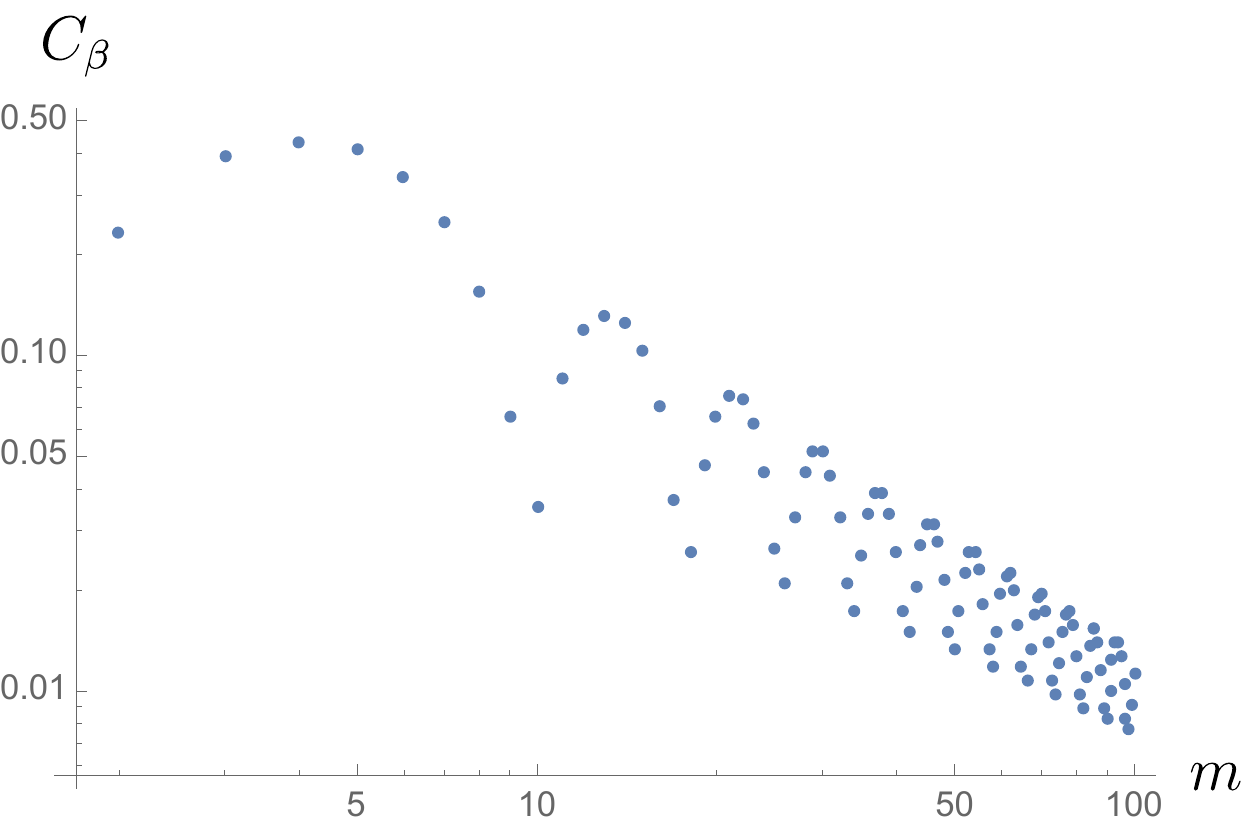}
    \caption{The absolute value of the late-time value of the OTOC for the unitary minimal models $\mathcal{M}(m,m+1)$ 
    % {\color{red}[p, p'?]} 
    is shown according to (\ref{mm_OTOC}). The OTOC asymptotes to zero as an oscillating power law ($m^{-1}$) at large $m$, showing enhanced scrambling. }
    \label{mm_OTOC_fig}
\end{figure}

% Now let's look at the case of minimal models.
The modular transformed partition function is
\bea
\mathcal{Z}\left(\frac{\tau}{1+2\tau},\bar \tau\right) =& \sum\limits_{r,s,\rho, \sigma }  (S \bar T^2 S)_{rs; \rho \sigma}  \Bigg[ K_{\lambda_{r,s}} (\tau) - K_{\lambda_{r,-s}} (\tau)  \Bigg]  \Bigg[ K_{\lambda_{\rho,\sigma}} (\bar \tau) - K_{\lambda_{\rho,-\sigma}} (\bar \tau)  \Bigg]
% \nonumber\\
% & \xlongequal{q \to 0} &  \sum\limits_{r,s,\rho, \sigma } (S \bar T^2 S)_{rs; \rho \sigma} q^{-\frac{1}{24}}    \bar q^{-\frac{1}{24}}   \Bigg[ q^{\frac{\lambda^2_{r,s}}{2N}} -q^{\frac{\lambda^2_{r,-s}}{2N}}  \Bigg]  \Bigg[ q^{\frac{\lambda^2_{\rho,\sigma}}{2N}} -q^{\frac{\lambda^2_{\rho,-\sigma}}{2N}}  \Bigg] 
% \nonumber\\
% &=& \sum\limits_{r,s,\rho, \sigma = 0}^{N-1} (S \bar T^2 S)_{rs; \rho \sigma} q^{-\frac{1}{24}}    \bar q^{-\frac{1}{24}}   \Bigg[ q^{\frac{pr - p's}{2N}} -q^{\frac{pr + p's}{2N}}  \Bigg]  \Bigg[ q^{\frac{p\rho - p' \sigma}{2N}} -q^{\frac{p\rho + p' \sigma}{2N}}  \Bigg]
\eea
We note that in the $q\rightarrow 0$ limit, this series is dominated by the $n=0$ term for $\lambda_{r,s}$
\begin{align}
    K_{\lambda_{r,s}}(\tau) \simeq  q^{\frac{\lambda_{r,s}^2}{2N} - \frac{1}{24}}.
\end{align}
% On the other hand, for $\lambda_{r,-s}$, the $n= -1$ term dominates, so
% \begin{align}
%     K_{\lambda_{r,-s}}(\tau) \simeq  q^{\frac{(\lambda_{r,-s}-2m(m+1))^2}{2N} - \frac{1}{24}}.
% \end{align}
Terms with $K_{\lambda_{r,-s}}$ are subdominant because their leading term $(n=-1)$ contains higher power of $q$.
According to B\'ezout's Lemma \cite{bezout2010theorie}, 
% {\color{blue}What is this?}
there is a unique pair of $(r_0, s_0)$ in the range of $1 \le r \le m-1, \ \ 1 \le s \le m$ that satisfies $(m+1)r_0 - ms_0 = 1$.
% Specifying to unitary minimal models where $p-p' = 1$,
In fact, we always have 
\begin{align}
    r_0 = 1,\quad s_0 =1
\end{align}
which is just the identity operator as predicted by \eqref{rcft_otoc_eq}. The partition function is then well approximated by
\begin{align}
    \mathcal{Z}\left(\frac{\tau}{1+2\tau},\bar \tau\right) &= (S\bar{T}^2 S)_{11;11}\left|q \right|^{-c/12},
\end{align}
so we confirm that
\begin{align}
    C_{\beta} \rightarrow {\frac{8}{m(m+1)}} \sum_{(r,s) \in E_{m+1,m}} \sin^2\left(\pi \frac{m+1}{m}r \right)\sin^2\left(\pi \frac{m}{m+1}s\right) e^{-4i\pi(h_{r,s}-c/24)}.
    \label{mm_OTOC}
\end{align}
At large $m$, this late-time value asymptotes as $m^{-1}$ to zero (see Fig.~\ref{mm_OTOC_fig}), showing that scrambling emerges as we enlarge the Hilbert space. This is consistent with the large $m$ analysis for the spectral form factor.

\subsubsection{TOMI}

\begin{figure}
    \centering
    \includegraphics[width = .48\textwidth]{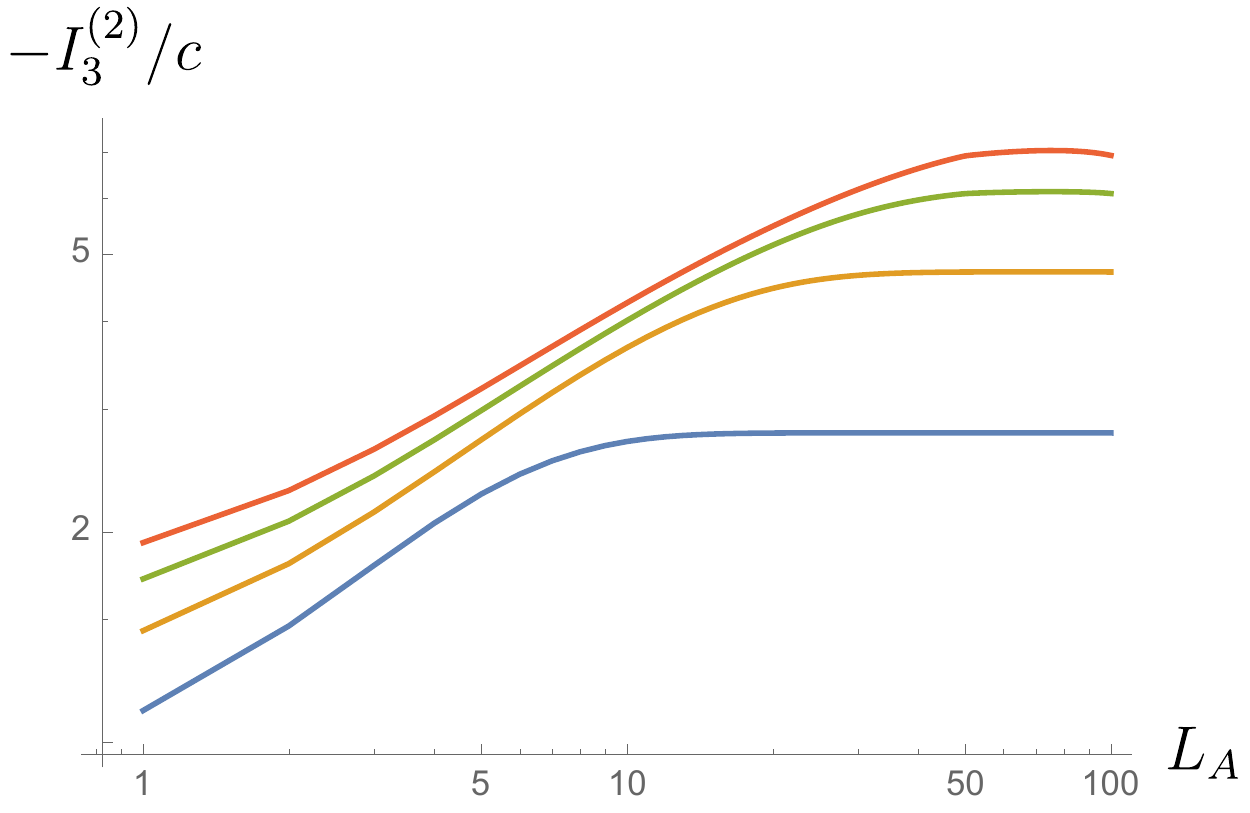}
    \includegraphics[width = .48\textwidth]{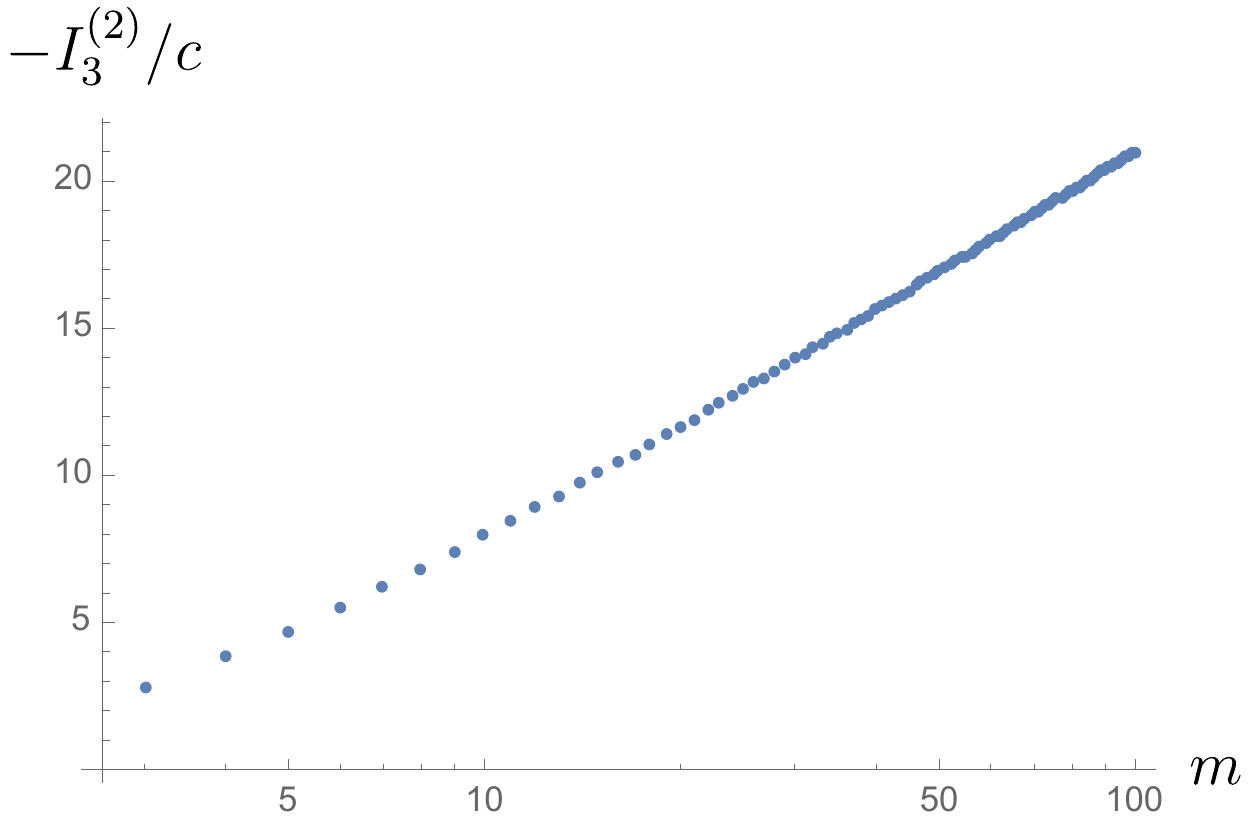}
    \caption{(left) The dependence on subsystem size $L_A$ of the saturation value of TOMI for $\mathcal{M}(m,m+1)$ with $m = \{ 3,5,7,9\}$ from bottom to top. The saturation to a constant occurs around $L_A/\epsilon \sim m^2$, before which, the growth is sublinear. (right) The saturation value in the scaling limit using (\ref{RCFT_I3}) shows logarithmic scaling with $m$. }
    \label{mm_scaling_TOMI}
\end{figure}

Because the minimal models are RCFTs, we can immediately apply our general formula for the late-time value of the TOMI \eqref{RCFT_I3}. Using \eqref{MM_S}, we find
\begin{align}
    I_3^{(2)} = \log \left[\frac{\sin^2\left(\pi\frac{m+1}{m}\right)\sin^2\left(\pi\frac{m}{m+1}\right)}{\sum_{h_{r,s}} \sin^2\left(\pi\frac{m+1}{m}r\right)\sin^2\left(\pi\frac{m}{m+1}s\right)} \right]
\end{align}
which is a constant that grows logarithmically with $m$ as shown in Fig.~\ref{mm_scaling_TOMI}. Even though in the scaling limit the late-time value is a constant that does not depend on the input system size, we are able to investigate the emergence of scrambling (delocalizing) behavior when considering finite-size effects. This is done by taking the size of $A$, $L_A/\epsilon$, to be of order $m$. In this case, sublinear scaling of the saturation value with subsystem size emerges, the onset of scrambling (Fig.~\ref{mm_scaling_TOMI} left).

\subsection{\texorpdfstring{$\mathfrak{su}(2)_k$}{1}}

Another tractable and interesting family of RCFTs is the $\mathfrak{su}(2)_k$ Wess-Zumino-Witten models whose Hilbert space decomposes as
\begin{align}
    \mathcal{H}_{WZW} = \bigoplus_{\lambda}R_{\lambda} \otimes R_{\lambda}.
\end{align}
In this diagonal sum, all representations appear once. The representations are labeled by $0 \leq \lambda \leq k$ which are equal to twice the spin. The central charge and conformal weights of the theory are given by
\begin{align}
    c = \frac{3k}{k+2},\quad h_{\lambda}^{(k)} = \frac{\lambda(\lambda+2)}{4(k+2)}. 
\end{align}
Given the above Hilbert space decomposition, the partition function is the diagonal modular invariant
\begin{align}
    \mathcal{Z}_{WZW} = \sum_{\lambda=0}^k \chi_{\lambda}^{(k)}(\tau)\bar \chi_{\lambda}^{(k)}(\bar{\tau}),
\end{align}
with characters given by
\begin{align}
    \chi_{\lambda}^{(k)}(q) = \frac{q^{(\lambda+1)^2/4(k+2)}}{\eta(q)^3}\sum_{n = -\infty}^{\infty}(\lambda + 1 + 2n(k+2))q^{n[\lambda+1+(k+2)n]}
\label{eq: su2kCharacters}
\end{align}

\subsubsection{SFF}
\label{su2_SFF}
\begin{figure}
    \centering
        \includegraphics[width=0.48\linewidth]{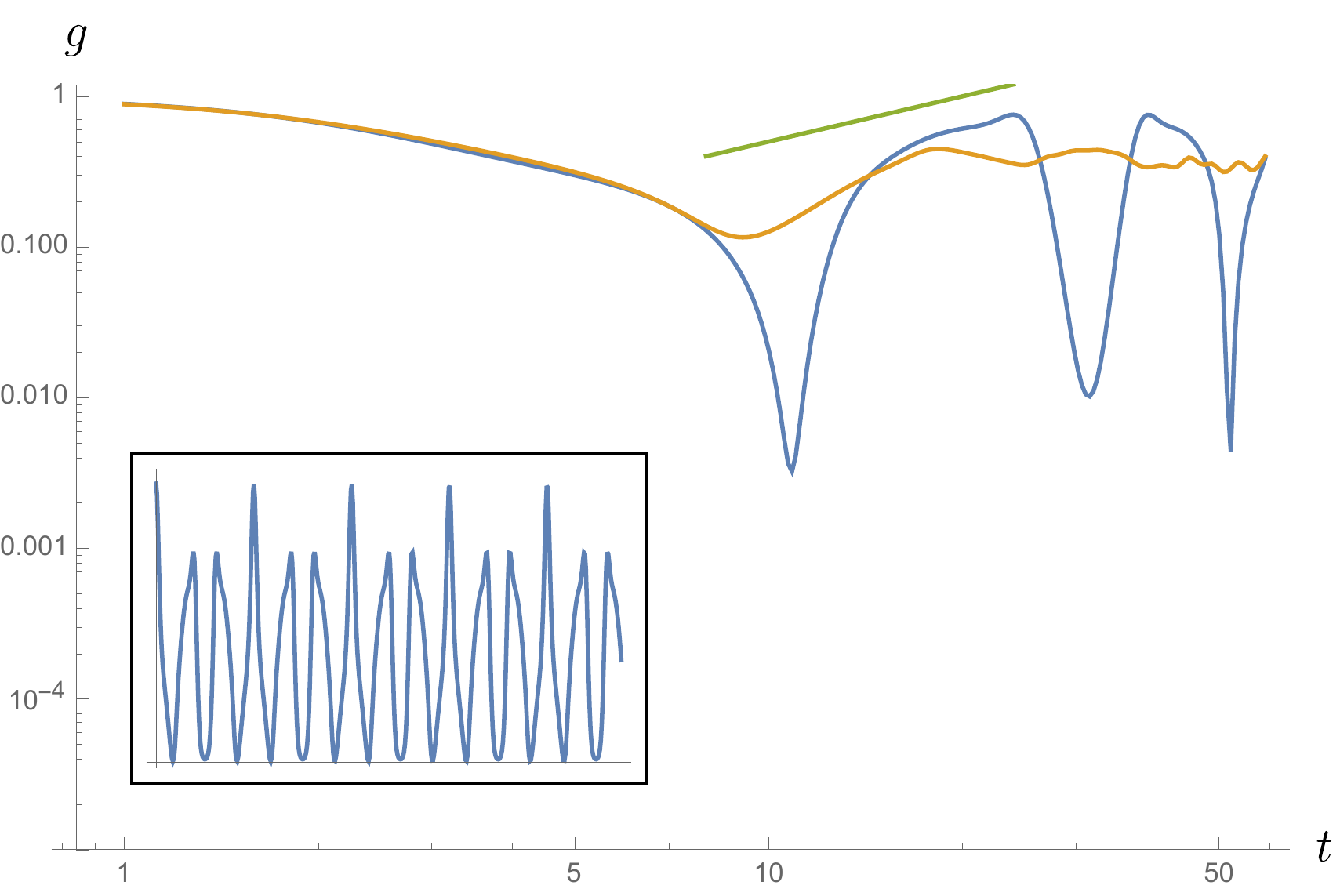}
        \includegraphics[width=0.48\linewidth]{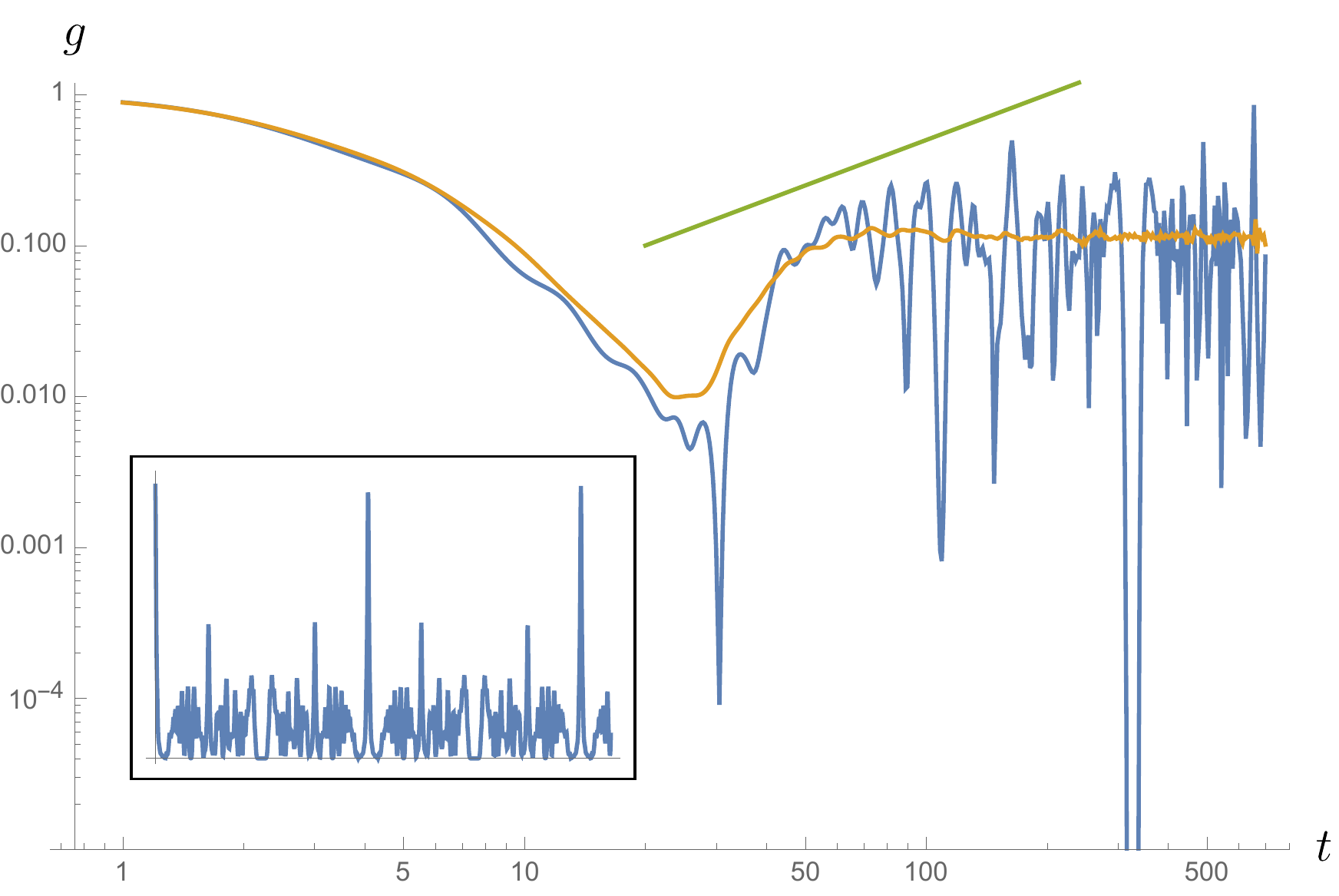}
        %  \caption{$k = 50$. Left: recurrence in non-log-log plot. Right: log-log plot for $t < t_{rec}$.}
    % \end{subfigure}
        \caption{Spectral form factor (blue) its progressive time-averaged version (yellow), and linear-in-$t$ curve (green, for comparison with the ramp) for $\mathfrak{su}(2)_k$ with $k=3$ (left) and $k=50$ (right). The dip, ramp, and plateau emerge, but the ramp is superlinear. The insets display the linearly scaled axes.}
    \label{fig: su(2)k_SFF}
\end{figure}

One can determine the recurrence time of the spectral form factor from the common denominator of the powers of $q = e^{-it - \beta}$ in~\eqref{eq: su2kCharacters}
\be
t_{rec} = 4 \pi (k+2).
\ee
The growing of the recurrence time with $k$ suggests that we may find emergent irrational structure in the spectral form factor if we take $k$ to be large enough.
Fig.~\ref{fig: su(2)k_SFF} demonstrates this emergence
% shows $\mathfrak{su}(2)_k$ spectral form factor in non-log-log plot, which shows recurrence, and in log-log plot, with dip-ramp-plateau structure for large $k$
after progressive time averaging. We note that the ``ramp" is superlinear,
reflecting a weaker rigidity of the spectrum. %hence does not describe strong spectral rigidity.

We can analytically approximate the initial decrease in the large-$k$ limit. For any fixed finite value of $\beta$, the sum over characters may be well approximated by an integral
\begin{align}
    \mathcal{Z} &= \sum_{\lambda = 0}^{k} \frac{\left(q \bar{q}\right)^{(\lambda+1)^2/4(k+2)}}{\left|\eta(q)\right|^6}(\lambda + 1) \rightarrow \frac{1}{\left|\eta(q)\right|^6}\int_0^{k} d\lambda \left(q\bar{q}\right)^{(\lambda+1)^2/4(k+2)}(\lambda +1 )
    \\
    &= \frac{2(2+k)\left(\left(q\bar{q}\right)^{\frac{(1+k)^2}{4(2+k)}}- \left(q\bar{q}\right)^{\frac{1}{4(2+k)} }\right)}{\left|\eta(q)\right|^6\log \left(q\bar{q}\right)}
    \simeq -\frac{2k  \left(q\bar{q}\right)^{\frac{1}{4(2+k)} }}{\left|\eta(q)\right|^6\log \left(q\bar{q}\right)}
\end{align}
This leads to $t^{-2}$ scaling at early times. Unfortunately, we are unable to fully understand the ramp from this analysis.

\subsubsection{OTOC}

\label{su2_OTOC}

\begin{figure}
    \centering
    \includegraphics[width = 9cm]{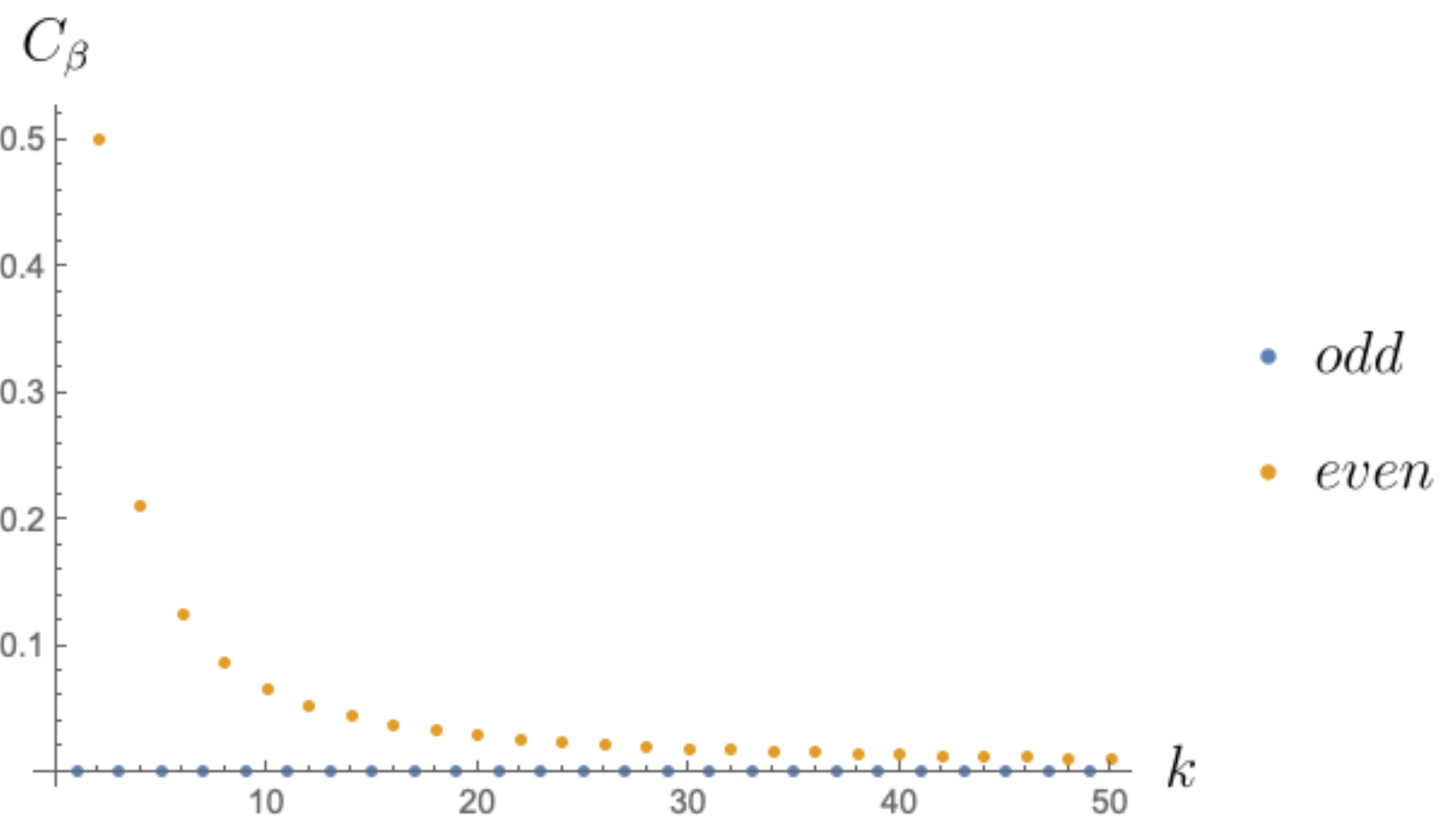}
    \caption{The absolute value of the OTOC at late times for the $\mathfrak{su}(2)_k$ WZW model (\ref{su2otoc}). Note the parity effect. 
    }
    \label{su2otocfig}
\end{figure}

For the OTOC, we are concerned with the $q\rightarrow 0$ limit where the $n = 0$ term in the characters is always dominates because the minimum at
\begin{align}
    \partial_n \left( n (\lambda + 1 + (k+2)n)\right) = 0
\end{align}
is always $\in (-1/2,1/2)$, so 
\begin{align}
    \chi_{\lambda}^{(k)}(q) \rightarrow \frac{q^{(\lambda+1)^2/4(k+2)}}{\eta(q)^3}(\lambda + 1).
\end{align}
Likewise, in the modular invariant, the $00$ term dominates
\begin{align}
    \mathcal{Z}\left(\frac{\tau}{1+2\tau}, \bar{\tau} \right) \rightarrow (S\bar{T}^2S)_{00}\frac{q^{1/4(k+2)}}{\eta(q)^3}\frac{\bar{q}^{1/4(k+2)}}{\eta(\bar{q})^3}
\end{align}
The modular matrices are known functions for $\mathfrak{su}(2)_k$ WZW models
\begin{align}
    \mathcal{T}_{\lambda \mu} = \delta_{\lambda \mu} e^{2\pi i m_{\lambda}}, \quad \mathcal{S}_{\lambda \mu} = \sqrt{\frac{2}{k+2}}\sin \left(\frac{\pi(\lambda+1)(\mu+1)}{(k+2)} \right) 
\end{align}
where $m_{\lambda}$ is the modular anomaly
\begin{align}
    m_{\lambda} = \frac{(\lambda+1)^2}{4(k+2)}.
\end{align}
The modular transformation needed is then 
\begin{align}
    (S\bar{T}^2S)_{00} =\frac{2}{k+2} \sum_{\lambda = 0}^k e^{-\frac{\pi i (\lambda +1)^2}{k+2}} \sin\left( \frac{\pi (\lambda+1)}{k+2}\right)
\end{align}
The late-time OTOC is
\begin{align}
    C_{\beta} \rightarrow 2^{-2k/(k+2)}\left| z \right|^{k/2(k+2)} (S\bar{T}^2S)_{00}\frac{q^{1/4(k+2)}}{\eta(q)^3}\frac{\bar{q}^{1/4(k+2)}}{\eta(\bar{q})^3}
\end{align}
% Because in this limit,
To leading order, the Dedekind eta function is just $q^{1/24}$, so we may confirm that 
\begin{align}
    C_{\beta}(t\rightarrow \infty) \simeq
    % 2^{(-2k+4)/(k+2)}\left| 1-z \right|^{-k/(k+2)} (S\bar{T}^2S)_{00}\rightarrow 
    % 2^{(-2k+4)/(k+2)} \frac{2}{k+2} 
    \sum_{\lambda = 0}^k e^{-\frac{\pi i (\lambda +1)^2}{k+2}} \sin\left( \frac{\pi (\lambda+1)}{k+2}\right)
    \label{su2otoc}
\end{align}
which has an interesting parity effect. For odd values of $k$, the late-time value is zero, while for even values, it is finite, though decreasing with $k$. This is shown in Fig.~\ref{su2otocfig}. We note that a similar parity effect was found for the compactified boson at rational radius in Ref.~\cite{2017PhRvD..96d6020C}.

One may worry that the trivial late-time value is merely an artifact of us taking the leading order term and higher order terms in the partition function will correct this. However, all higher order terms contain additional positive powers of $q$ that will exponential decay to zero.

\subsubsection{TOMI}
\label{su2_TOMI}

We investigate the scaling of TOMI as we increase the level. We can use our general formula for RCFT (\ref{RCFT_I3}) and explicitly evaluate the sum to find
\begin{align}
    I_3 = -\log \left[\frac{ \csc\left(\frac{\pi}{2+k} \right)^2\left(3+2k-\csc\left( \frac{\pi}{2+k}\right)\sin\left(\frac{(3+2k)\pi}{2+k}\right)\right) }{4}\right].
    \label{su2_TOMI_eq}
\end{align}
In the large-$k$ limit, this scales as
\begin{align}
    I_3 \rightarrow - \log \left[\frac{k^3}{2\pi^2} \right]
\end{align}
% We find that the absolute value at late-times (per central charge) increases as a power law that transitions from $\sqrt{k}$ to $k^{1/6}$ (see Fig.~\ref{su2_TOMI_varyk}) \textcolor{blue}{[This transition is potentially occurring when $k \sim L_A/\epsilon$]}.

\begin{figure}
    \centering
    \includegraphics[width = .48\textwidth]{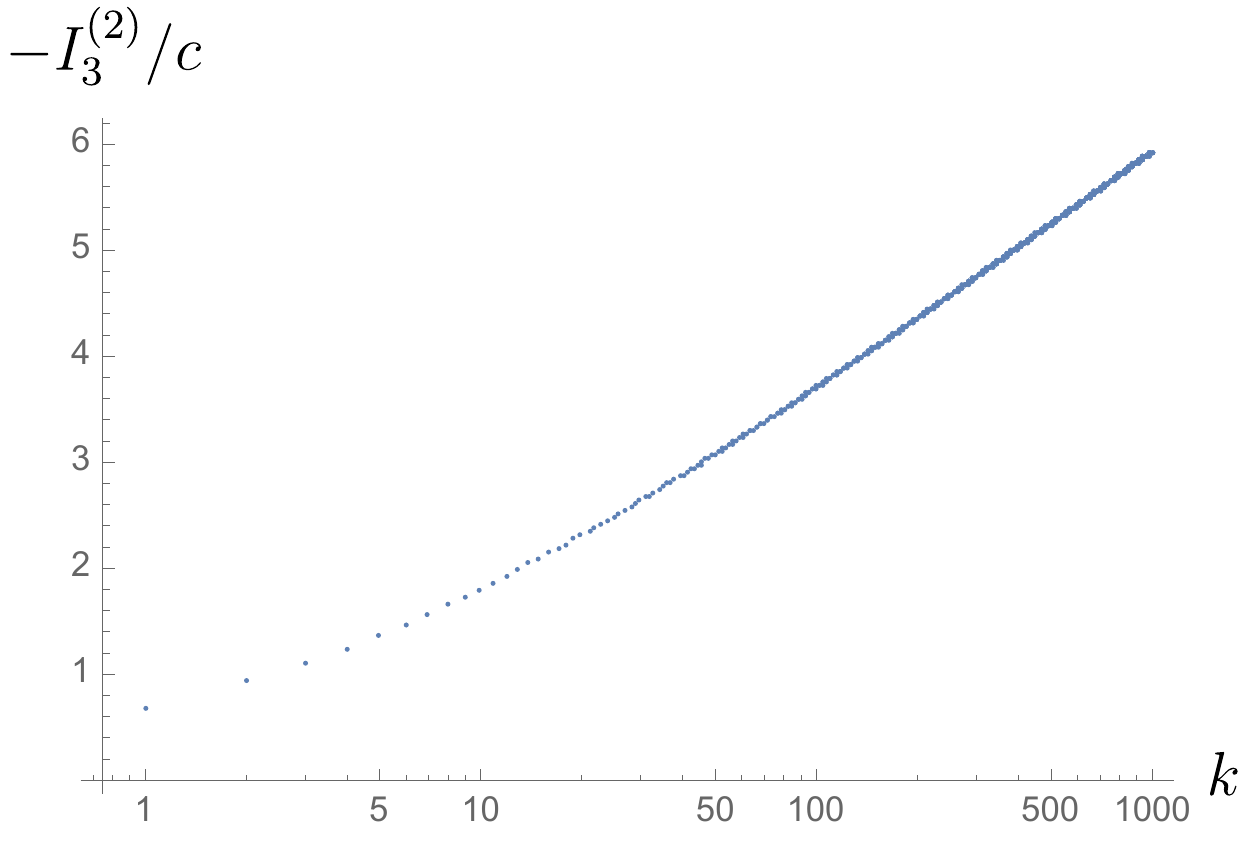}
    \includegraphics[width = .48\textwidth]{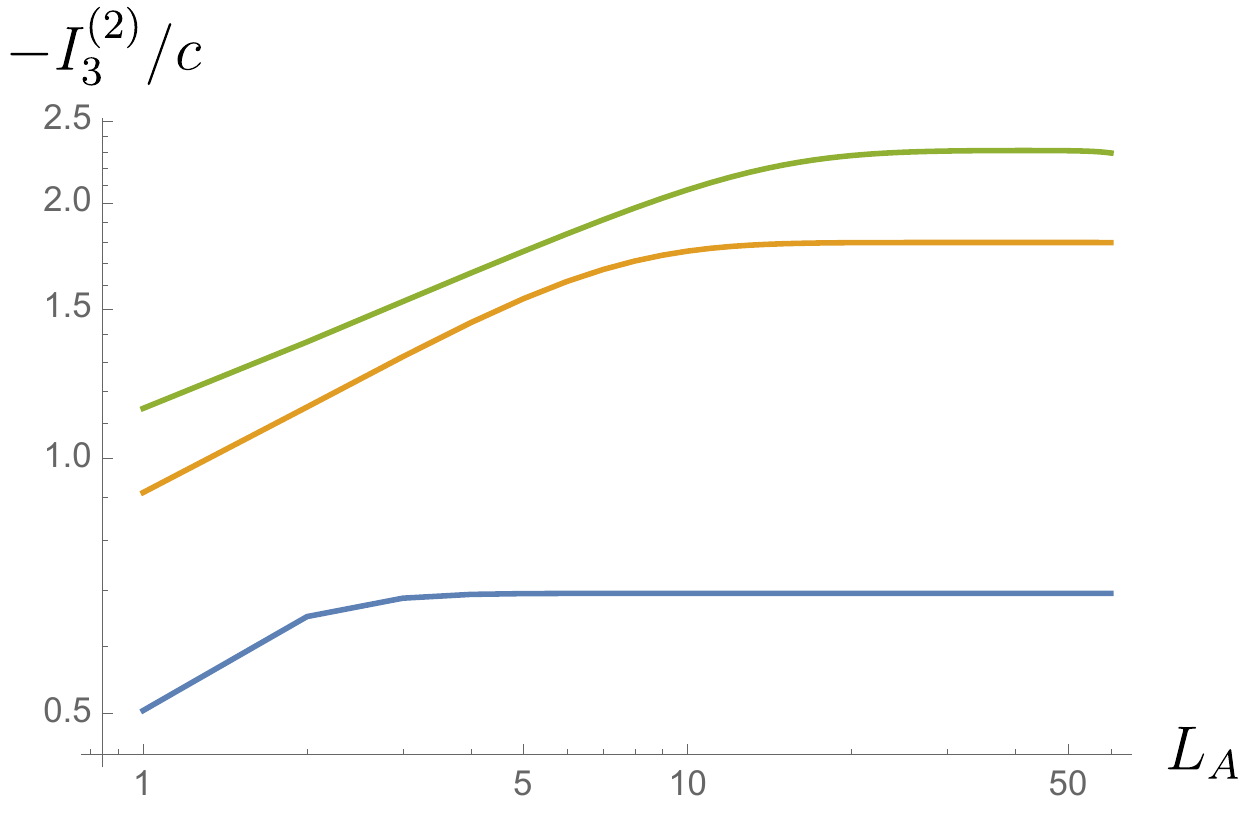}
    \caption{Left: (blue) the late-time value of TOMI in the scaling limit grows logarithmically with $k$ \eqref{su2_TOMI_eq}. Right: $k = \{1,10,20 \}$ from bottom to top. Saturation occurs around $k\sim L_A/\epsilon$, before which, there is sublinear growth. When subsystem size overcomes the size of the local Hilbert space size, the TOMI saturates. }
    \label{su2_TOMI_fig}
\end{figure}

As for the minimal models, it is also of interest to understand how TOMI depends on the subsystem size $L_A$ before taking the scaling limit. There is sublinear scaling with system size when $L_A/\epsilon$ is less than the level (Fig.~\ref{su2_TOMI_fig} right). Again, this shows the interplay between local Hilbert space size and subsystem size in the onset of scrambling.
% We find that the saturation value is decreasing for $L_A < \epsilon$ but becomes constant for sufficiently large subsystems $L_A/\epsilon \sim k$. This is shown in Fig.~\ref{su2_TOMI_scale}

% \begin{figure}
%     \centering

%     \caption{}
%     \label{su2_TOMI_scale}
% \end{figure}

% \subsubsection{Large k limit}

% In the large level limit, the characters become
% \begin{align}
%     \chi_{\lambda}^{(k)}|_{k\rightarrow \infty} = \frac{q^{(\lambda+1)^2/4(k+2)}}{\eta(q)^3}.
% \end{align}
% Note that this is the same asymptotic formula seen above for the small $q$ limit. It was then found in \cite{2018NuPhB.934..498F} that the partition function becomes proportional to three free bosons on $\mathbb{R}$
% \begin{align}
%     \mathcal{Z}_{WZW}|_{k\rightarrow \infty} = k^{3/2}\left(\frac{1}{\sqrt{\Im (\tau)}}\frac{1}{\left|\eta(q) \right|^2} \right)^3.
% \end{align}
% Only the primaries with $\lambda \propto \sqrt{k}$ contribute to this modular invariant. 
% % Because of the equivalence to three free bosons, we conclude that $\mathfrak{su}(2)_k$ WZW models at large level do not scramble quantum information.

\section{Irrational CFT (\texorpdfstring{$c<\infty$}{2})}

\label{sec_ICFT}

Unfortunately, our abilities to characterize irrational conformal field theories are somewhat limited compared to rational CFTs. This is largely due to the absence of explicit constructions of modular invariant partition functions of such theories. We will first make some general arguments based solely on the partition function and the number of conserved currents. Then, we choose the tractable and interesting example of the compactified boson at irrational radius. In the next section, we study CFTs at large central charge possessing a weakly-coupled gravity dual. These will serve as the extreme limiting behavior of irrational CFTs.

\paragraph{SFF}

For generic irrational CFTs, there is not a lot we can say for sure about the spectral form factor. One thing that distinguishes it from the RCFTs is the infinite number of characters in the partition function. This leads to the recurrence time of the spectral form factor to generically be infinite. Moreover, there is an expectation that generic irrational CFTs will possess random matrix theory statistics; thus, they should display a dip, linear ramp, and plateau. We currently do not have more quantitative generic results to report.

\paragraph{OTOC}
If we restrict ourselves to theories with central charge larger than 1 and no conserved currents beyond the stress tensor, general statements may be made about the late-time value of the OTOC for twist-fields.
% We can demonstrate this by reviewing the generic result for the late-time behavior of conformal field theories with central charge larger than one and no extended symmetry algebra (note that the compact boson has a $U(1)$ current).
Without specifying any particular theory, the OTOC may be computed by taking advantage of the Regge limit of the dominant conformal block in the four-point function \cite{2019arXiv190502191K}
\begin{align}
    C_{\beta} (x,t) \rightarrow 
    \begin{cases}
        e^{-h_{2 \alpha_W} \frac{2\pi t}{\beta} }, & \alpha_W < \frac{Q}{4}
        \\
        e^{-h_{2 \alpha_V} \frac{2\pi t}{\beta} }, & \alpha_V < \frac{Q}{4}
        \\
        e^{-\frac{Q^2}{4}\frac{2\pi t}{\beta} }t^{-\frac{3}{2}}, & \alpha_W, \alpha_V > \frac{Q}{4}
    \end{cases}
    \label{pure_otoc}
\end{align}
where we are using Liouville coordinates
\begin{align}
    c = 1 + 6Q^2, \quad Q = b+b^{-1}, \quad h_{\alpha} = \alpha(Q-\alpha).
\end{align}
The exponential decay is therefore a generic feature of irrational CFTs regardless of the operators used. For twist fields of the $\mathbb{Z}_2$ orbifold, we have
\begin{align}
    \alpha_{\sigma_2} = \frac{1}{2}\left( Q-\frac{1}{2}\sqrt{-1-2Q^2}\right),
\end{align}
the real part of which is always greater than $Q/4$ as long as $c> 1$. Thus, we generically find that the OTOC decays to zero at late times as\footnote{The power of $-3/2$ is related to the universal behavior of Virasoro blocks at late times numerically observed in Ref.~\cite{Chen2017}.}
\begin{align}
    C_{\beta} \rightarrow  e^{-\frac{\pi (c-1) t}{12\beta} }t^{-\frac{3}{2}}.
    \label{ICFT_OTOC_eq}
\end{align}

\paragraph{TOMI}
We resume our discussion from Sec.~\ref{sec_RCFT} regarding general statements about operator entanglement in irrational CFTs. Some of the arguments for RCFT will not follow through to irrational CFTs due to the infinite number of characters and nontrivial modular transformations. However, we can make generic bounds. For the BOMI, we have
\begin{align}
   0 \leq  I_{AB_1}, I_{AB_2} \leq \frac{\pi l (2c_{currents}+c)}{24 \epsilon }+ \frac{c_{currents}-c}{3}\log 2+ \log S_{00}.
\end{align}
The lower bound is enforced by positivity of mutual information and represents maximal scrambling. The upper bound is nontrivial and shows that decreasing the number of conserved currents increases the information   scrambling. The quasi-particle picture is recovered for $c_{currents } = c$. We may also bound the TOMI between the quasi-particle picture and the linear result (\ref{I3_RCFT})
\begin{align}
    0 \geq I_3 \geq -\frac{\pi cl }{4\epsilon } + \log \frac{S_{00}^2}{\sum_h S_{h0}^2}.
    \label{I3_bound}
\end{align}
% This hints that linear scaling of TOMI may be somewhat generic in irrational CFTs.
% In fact, this bound is tight, as we show in the following section. 
% We note that holographic CFTs saturate this bound (up to the constant piece), providing further evidence that holographic CFTs delocalize information the most effectively.

% Again, we note that the compactified boson at irrational radius neither grows linearly or is a constant, representing a transition between rational and irrational theories.

% Keeping this in mind, we proceed.

\subsection{Compactified boson}

% \textcolor{red}{discuss intermediate OTOC of Ref.~\cite{2017PhRvD..96d6020C}.}
Though the compactified boson has an infinite number of representations, we stress that it is not a a typical representative of the scrambling and chaos believed to exist in irrational CFTs. In particular, it is different than the CFTs discussed above 
% due to its $U(1)$ current that renders 
because $c_{currents} = c =1$. Nevertheless, the compactified boson is an interesting model to study because we can tune the rationality of the theory by the rationality of the square of the compactification radius $\eta$\footnote{We recommend the reader be careful as we have used the same symbol here as we are using for the Dedekind eta functions that are used throughout.}.
% It was demonstrated in Ref.~\cite{2018arXiv181200013N} that theories at irrational radii showed novel signatures of scrambling involving the tripartite operator mutual information scaling as $\log L_A/\epsilon$. We would like to see how/if this signature shows up in the spectral form factor. 
The partition function for the compact boson is
\begin{align}
    \mathcal{Z}_{\eta}(\tau, \bar{\tau}) = \frac{\Theta(0|T)}{\left| \eta(\tau)\right|^2}
\end{align}
where 
\begin{align}
    \Theta(0|T) &= \sum_{\mu, \nu \in \mathbb{Z}} \exp\left[\frac{\pi i \tau}{2}\left(\nu \sqrt{\eta} + \frac{\mu}{\sqrt{\eta}}\right)^2 \right]\exp\left[-\frac{\pi i \bar{\tau}}{2}\left(\nu \sqrt{\eta} - \frac{\mu}{\sqrt{\eta}}\right)^2 \right]
    \label{siegel_theta} \nonumber\\
    &= \sum_{\mu, \nu \in \mathbb{Z}} \exp \left[ i \pi \left(\nu \ \  \mu \right) T  \left( \begin{array}{ccc}   
\nu  \\
\mu \end{array} \right) \right]
\end{align}
and the modular matrix is
\begin{align}
    T  = \begin{pmatrix} 
    \eta \frac{\tau -\bar{\tau}}{2}& \frac{\tau + \bar{\tau}}{2}\\
    \frac{\tau + \bar{\tau}}{2} & \frac{1}{\eta} \frac{\tau -\bar{\tau}}{2}
    \end{pmatrix}.
\end{align}

\subsubsection{SFF}
\label{CB_SFF}
\begin{figure}
    % \begin{subfigure}
    \centering
    \includegraphics[width=0.48\linewidth]{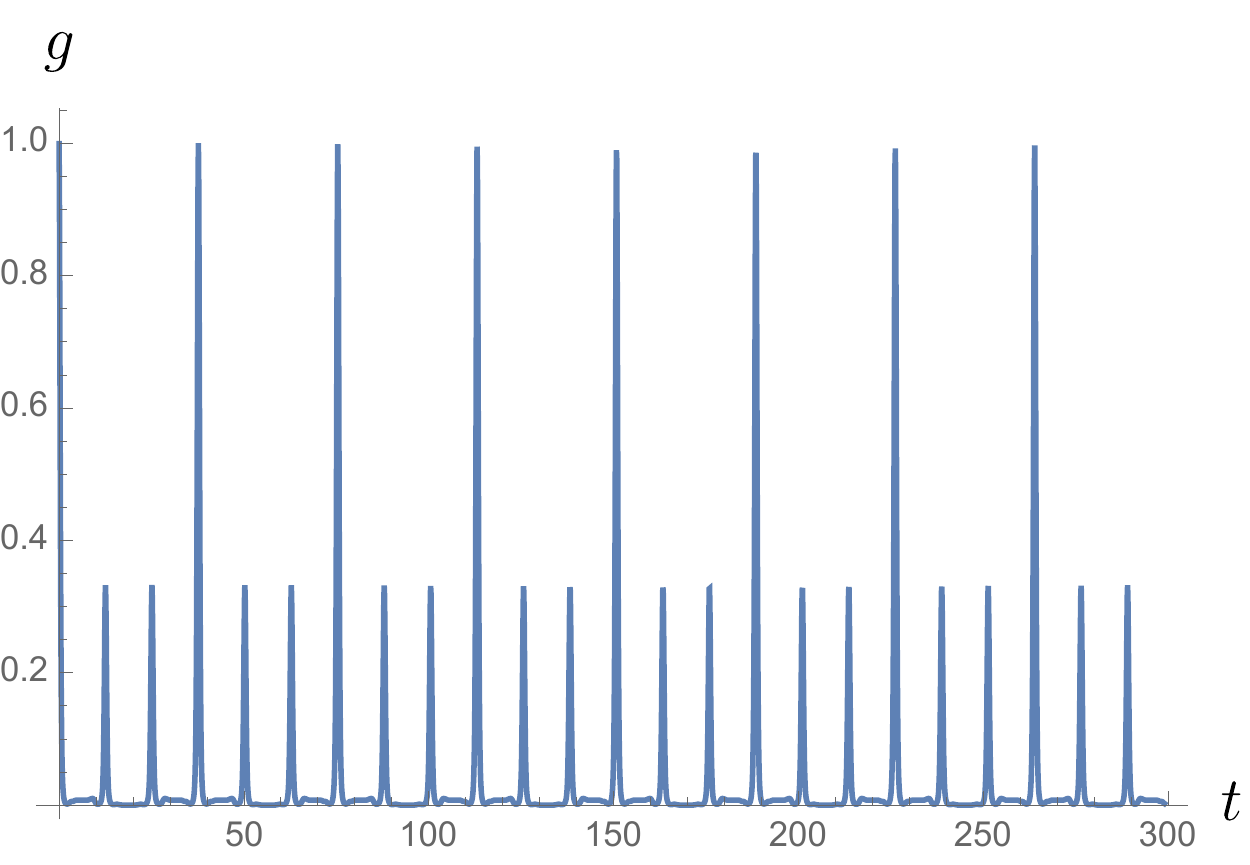}
    \includegraphics[width=0.48\linewidth]{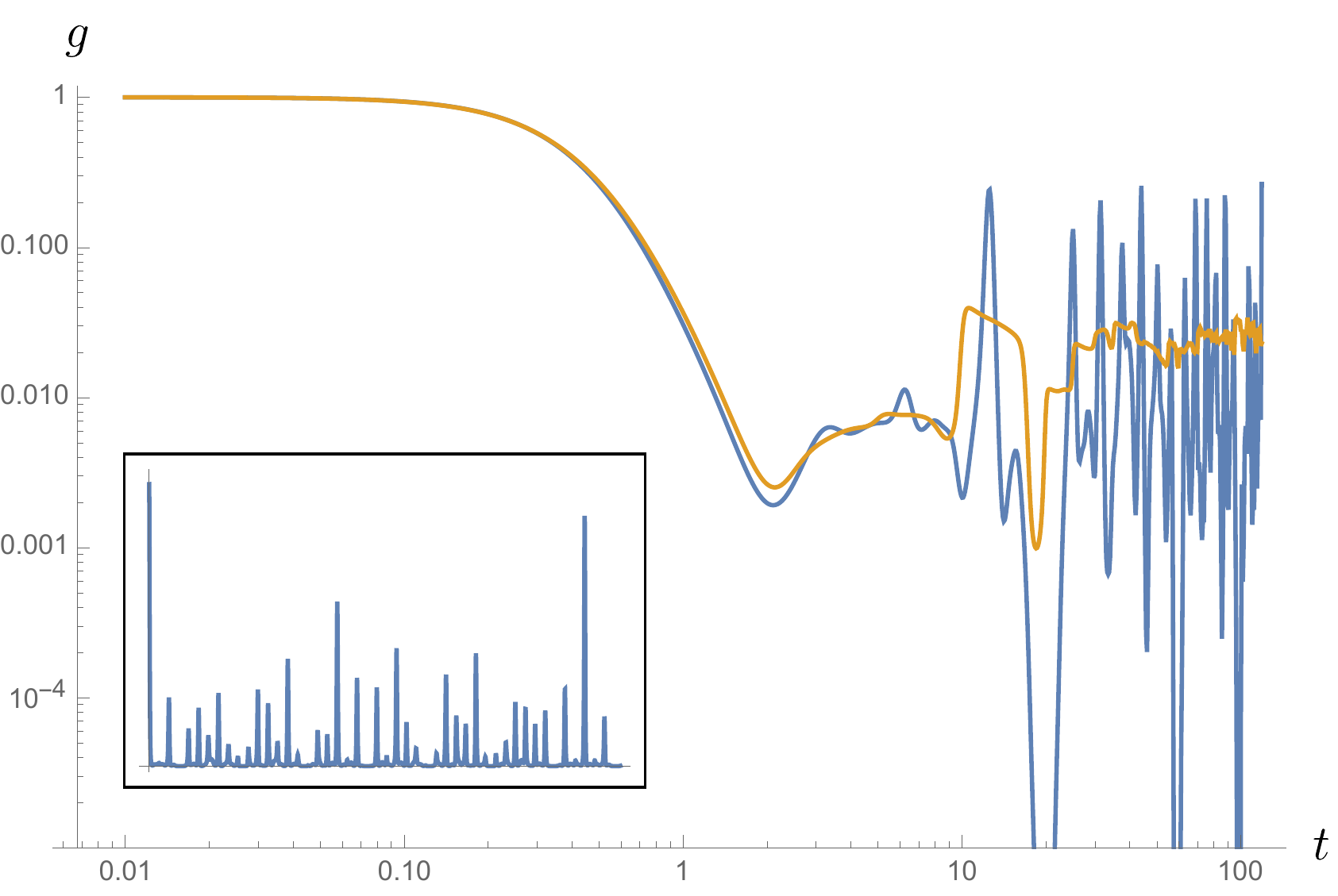}
        % \caption{$\eta = \pi$, $t_{rec} = \infty$, $\alpha = 0.9$. Left: recurrence in non-log-log plot. Right: log-log plot for $t < t_{rec}$.}
    % \end{subfigure}
    \caption{Spectral form factor (blue) and its progressive time-averaged version (yellow) for (left) rational ($\eta = 3$) and (right) irrational ($\eta = \pi$) compactified boson theories. For progressive time averaging, we use $\alpha = 0.9$. Clearly, the spectral form factor has finite recurrence for the rational radius. It has infinite recurrence time for irrational radius. The inset shows linear axes scaling for the irrational case.}
    \label{fig: CB_SFF}
\end{figure}

% The spectral form factor is defined as 
% \begin{align}
%     g(\beta, t) \equiv \left| \mathcal{Z}(\beta + i t)\right|^2 = \mathcal{Z}(\beta + i t) \mathcal{Z}(\beta - i t) 
% \end{align}
% where the real time $t$ is related to $\tau$ through analytic continuation
% \be
% \tau = \frac{i}{2\pi}(\beta -it) = - \bar \tau.
% \ee
We compare the spectral form factor for the compact boson for rational and irrational radii in Fig.~\ref{fig: CB_SFF}. The rational case clearly shows periodic behavior, thus a finite recurrence time, while the irrational case is less regular at late times, giving $t_{rec} = \infty$ as one would hope. Even so, the irrational compact boson does not show any obvious ramp structure, though it has a dip and plateau. As these are numerical observations, one may worry that we are unable to properly observe the irrational radius because computers only know rational numbers. However, we note that this is merely another example of rational theories appearing irrational in certain limits, analogous to the large $m$ and $k$ limits from Sec.~\ref{sec_RCFT}. While the numbers used must be rational, the relatively prime numerators and denominators are large, mimicking an irrational theory.

% will only keep one figure (with two subfigures) here, for rational and irrational radii. Each figure contains SFF and PAverage SFF. Need longer time, e.g. t = 600

% OAverage SFF and explanations of why progressive time average practically works will be placed in Appendix.

\subsubsection{OTOC}
\label{CB_OTOC}

The OTOC for twist fields for the compactified boson CFT has been sufficiently studied in Ref.~\cite{2017PhRvD..96d6020C}. This is where the idea for using twist fields as the operators in the OTOC originated. For completeness, we reproduce the results. For rational radius ($\eta = p/p'$), the OTOC has a parity effect similar to $\mathfrak{su}(2)_k$
\begin{align}
    C_{\beta}\rightarrow \begin{cases}
    \frac{1}{p p'}, & pp'\in 2\mathbb{Z}
    \\
    0 ,& pp' \in 2\mathbb{Z}+1
    \end{cases}.
\end{align}
For irrational $\eta$
\begin{align}
    C_{\beta} \rightarrow -\frac{\pi}{2 \log\left( -\frac{\epsilon_{12}*\epsilon_{34}}{16}e^{-\frac{2\pi(t-x)}{\beta}}\right)}
\end{align}
which describes a polynomial decay to 0. This shows the intermediate scrambling behavior of the compact boson at irrational radius. It is not exponential like generic irrational CFTs, but it is also not a constant like RCFTs. This is similar to the spectral form factor which had infinite recurrence time, but no dip-ramp-plateau structure.

\subsubsection{TOMI}
\label{CB_TOMI}
The operator entanglement for the compactified boson at both rational and irrational radii has been extensively studied in Ref.~\cite{2018arXiv181200013N}. Again, we reproduce the results for completeness, now with an understanding from \eqref{RCFT_I3} where the rational result originates from. For rational $\eta$
\begin{align}
    I_3^{(2)} = \begin{cases}
    -2 \log pp', & 2pp', \eta \ll L_A/\epsilon 
    \\
    -2 \log L_A/\epsilon, & \eta\ll L_A/\epsilon \ll 2pp'
    \\
    -\log \eta L_A/\epsilon, & L_A/\epsilon \ll \eta, 2pp'
    \end{cases}
\end{align}
This clearly shows the emergence of irrational behavior when the local Hilbert space dimension is larger than the subsystem, just like minimal models and WZW models. We expect this to be a generic features of quantum systems. For the irrational $\eta$
\begin{align}
        I_3^{(2)} = \begin{cases}
    -2 \log L_A/\epsilon, & \eta\ll L_A/\epsilon
    \\
    -\log \eta L_A/\epsilon, & L_A/\epsilon \ll \eta
    \end{cases}.
\end{align}

\section{Holographic CFT}

\label{sec_HCFT}

We conclude with analysis of conformal field theories at large central charge and possessing a semiclassical holographic dual. Understanding this class of CFTs has recently been been a central motivation in studying quantum many-body chaos. It is an open question as to which criteria are sufficient for a conformal field theories to admit a holographic dual describing smooth Einstein gravity (beyond the large $N$ constraint \cite{1999IJTP...38.1113M}). Significant progress has been made on constraining the spectrum of candidate CFTs by considering the 3d gravitational thermal partition function \cite{2014JHEP...09..118H} where it was determined that holographic 2D CFTs must have suitably sparse low-lying spectra. An intriguing additional criterion was proposed in Ref.~\cite{2016JHEP...08..106M} stating that the dual CFTs must be \textit{maximally chaotic} in that they saturate the chaos bound
\begin{align}
    \lambda_L \leq \frac{2\pi}{\beta}.
\end{align}
This is in the same line of thought as thinking of black holes as natures fastest scramblers \cite{2008JHEP...10..065S}. It is then fruitful to investigate what the torus partition function tells us about holographic CFTs. The beauty of holography is that we do not need to know the precise microscopic theory on the boundary to have certain analytic control of the partition function. This is due to the basic entry in the AdS/CFT dictionary that the partition functions are equal
\begin{align}
    \mathcal{Z}_{AdS} = \mathcal{Z}_{CFT}.
\end{align}
Thus, the CFT partition function on the torus is equal to the gravity partition function on the solid torus. This partition function is also parametrized by the modular parameter and has been studied extensively in the literature. Famously, the Hawking-Page transition describes how the dominant bulk saddle point switches at $\tau = \frac{2\pi}{\beta}$ from ``filling in" the torus with a thermal gas with periodic Euclidean time to the BTZ black hole geometry \cite{Hawking1983,1992PhRvL..69.1849B}. From the CFT perspective, this corresponds to a modular S transformation of the vacuum character
\begin{align}
    \chi_0 = e^{-\frac{i\pi c \tau}{12} }.
\end{align}
Each bulk saddle is not modular invariant on its own, but one way to construct a modular invariant function is to compute a Poincar\'e series, summing over all modular transformations of the vacuum character corresponding to thermal $AdS$
\begin{align}
    \mathcal{Z}\left(\tau, \bar{\tau} \right) = \sum_{PSL(2, \mathbb{Z})} \chi_0(\gamma(\tau)) \bar{\chi}_0(\gamma(\bar{\tau})),
\end{align}
where $PSL(2, \mathbb{Z})$ takes values in the modular group 
\begin{align}
    \gamma(\tau) = \frac{a \tau + b}{ c\tau + d }, \quad ad - bc = 1.
\end{align}
These sums, for example, have been of some interest in the search for a theory of pure quantum gravity in three dimensions, i.e. a dynamical metric and no additional quantum fields \cite{2010JHEP...02..029M,2015JHEP...02..080K}. Due to the large-c nature of the characters, only one character will dominate the partition function at a given ``time."
As it turns out \cite{2015JHEP...09..110A}, the Hawking-Page transitions is the only possible phase transition when the cross-ratios are between 0 and 1. This corresponds to purely imaginary modular parameters. In this case, the partition function (at tree level) is an optimization
\begin{align}
    \mathcal{Z}(\tau , \bar{\tau}) = \exp\left[\frac{  \pi c}{12} \max\left[i\tau^{-1}-i\bar{\tau}^{-1} ,i\bar{\tau}-i\tau \right] \right].
\end{align}

\subsection{SFF}
\label{holo_SFF}

\begin{figure}
    \centering
    \includegraphics[height = 4.5cm]{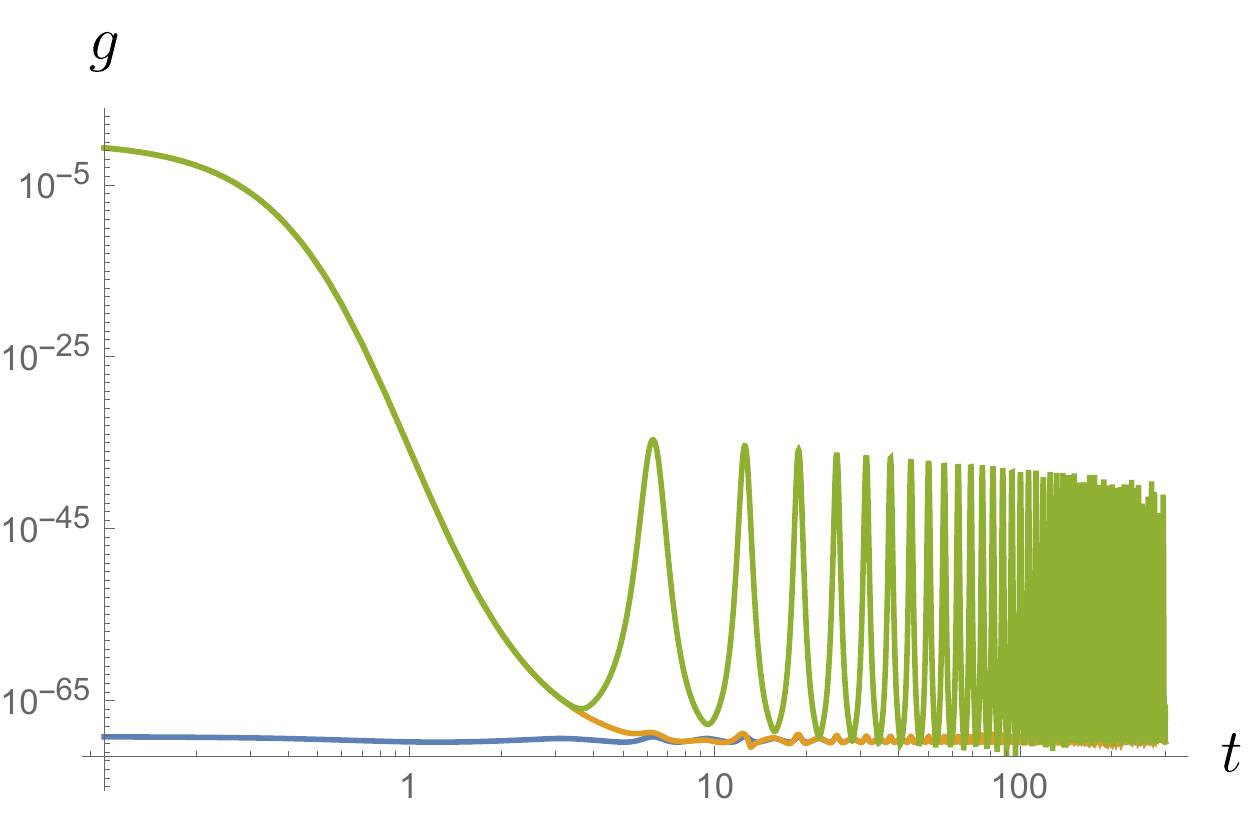}
    \caption{The (normalized) spectral form factor for thermal AdS (blue), the BTZ black hole (yellow), and many saddles (green). The ``plateau" decays at large times, showing information loss. We take $c = 25$ and $\beta = 1$. 
    % {\color{red}Will progressive time averaging help this?}
    }
    \label{SFF_holo_saddles}
\end{figure}

The spectral form factor for holographic CFTs is particularly interesting because, in theory, it should probe the fine-grained structure of the spectrum of quantum gravity. In particular, a plateau should emerge due to the expected discreteness of the spectrum. This is related to one form of the information paradox formalized by Maldacena \cite{2003JHEP...04..021M}. In Ref.~\cite{2017JHEP...08..075D}, the spectral form factor for $AdS_3$ gravity was carefully studied and it was shown that one must consider saddle points well beyond just Thermal $AdS$ and the BTZ black hole. These become important because the modular parameter in not purely imaginary in the analytic continuation for the spectral form factor \eqref{SFF_analytic cont}.
In particular, the spectral form factor does not display the characteristic ramp and plateau if one only considers these leading saddle for the partition function \cite{2017JHEP...08..075D}. Rather, only the universal early-time decay is found. Each saddle in the Poincar\'e series enjoys the spotlight momentarily; even so, the plateau decays away (Fig.~\ref{SFF_holo_saddles}), so ``information restoration" from this perspective is beyond present understanding.

\subsection{OTOC}
\label{holo_OTOC}

For the OTOC, the cross-ratios may be complex, so, in principle, the subleading saddles may become important. However, we find that this is not the case and the Thermal $AdS$ character dominates the partition function at all relevant times\footnote{This auxiliary Thermal $AdS$ partition function needed to compute the OTOC should not be confused with the physical partition function of the state which is the BTZ black hole because we are in the high-temperature regime.}. At late times,
\begin{align}
    \mathcal{Z}_{TAdS}\left( \frac{\tau}{1+2\tau}, \bar{\tau} \right) &= \exp \left[-\frac{\pi c}{12} \left(i\frac{\tau}{1+2\tau} -i\bar{\tau} \right)\right]
    % \\
    % &\simeq \exp \left[\frac{ c}{12} \left(\frac{\log\left(\frac{16}{z} \right)}{\left(1+\frac{2i}{\pi} \log\left( \frac{16}{z}\right)\right)} +{\log\left(\frac{16}{\bar{z}} \right)} \right)\right]
    \nonumber
    \\
    &\simeq \exp \left[\frac{ c}{12} \left(\frac{\alpha + \frac{2\pi}{\beta} (t-x)}{\left(1+\frac{2i}{\pi}\left(\alpha + \frac{2\pi}{\beta} (t-x)\right)\right)} +\alpha + \frac{2\pi}{\beta} (t+x)\right)\right],
\end{align}
where we have defined
\begin{align}
    \alpha \equiv \log \left(\frac{16}{-\epsilon^*_{12}\epsilon_{34}} \right).
\end{align}
Thus, the late-time behavior of the OTOC is
\begin{align}
    C_{\beta} &= 2^{-2c/3}\left| 1-z \right|^{-c/12} \left| z \right|^{c/6} \exp \left[\frac{ c}{12} \left(\frac{\alpha + \frac{2\pi}{\beta} (t-x)}{\left(1+\frac{2i}{\pi}\left(\alpha + \frac{2\pi}{\beta} (t-x)\right)\right)} +\alpha + \frac{2\pi}{\beta} (t+x)\right)\right]
   \nonumber \\
    &= 2^{-2c/3}\left( (1+e^{-(2\pi/\beta)(t-x)}\epsilon^*_{12}\epsilon_{34})(1+e^{-(2\pi/\beta)(t+x)}\epsilon^*_{12}\epsilon_{34}) \right)^{-c/24} \left( e^{-(2\pi/\beta)(t)}\epsilon^*_{12}\epsilon_{34}\right)^{c/6} 
   \nonumber \\ 
    &\times \exp \left[\frac{ c}{12} \left(\frac{\alpha + \frac{2\pi}{\beta} (t-x)}{\left(1+\frac{2i}{\pi}\left(\alpha + \frac{2\pi}{\beta} (t-x)\right)\right)} +\alpha + \frac{2\pi}{\beta} (t+x)\right)\right]
\end{align}
sending the regulators to zeros and taking $t \gg x$, we find the scaling
\begin{align}
    % &\simeq 2^{-2c/3} \left( e^{-(2\pi/\beta)(t)}\epsilon^*_{12}\epsilon_{34}\right)^{c/6} \left(\frac{16}{-\epsilon^*_{12}\epsilon_{34}}\right)^{\frac{  c}{12} }e^{2\pi c t/12\beta}
    % \\
    C_{\beta}&\propto e^{-2\pi c t/12\beta}
    \label{HHHH_OTOC}
\end{align}
This is consistent with the general result for irrational CFTs \eqref{ICFT_OTOC_eq} once we take the large-c limit and recall that we are in the $\mathbb{Z}_2$ orbifold theory, so the central charge is doubled. Interestingly, the exponent is different from the late-time behavior of the OTOC of heavy and light operators which scales as 
\begin{align}
  C_{\beta}  &\propto e^{-2\pi \Delta_{\mathcal{O}} t/\beta}.
  \label{HHLL_OTOC}
\end{align}
This follows from large-$c$ analysis of the first two lines of \eqref{ICFT_OTOC_eq}.
The distinction between \eqref{HHHH_OTOC} and \eqref{HHLL_OTOC} may be attributed to the fact that all four twist operators scale linearly with the central charge.

\subsection{TOMI}
\label{holo_TOMI}
\begin{figure}
    \centering
    \includegraphics[height = 4.5cm]{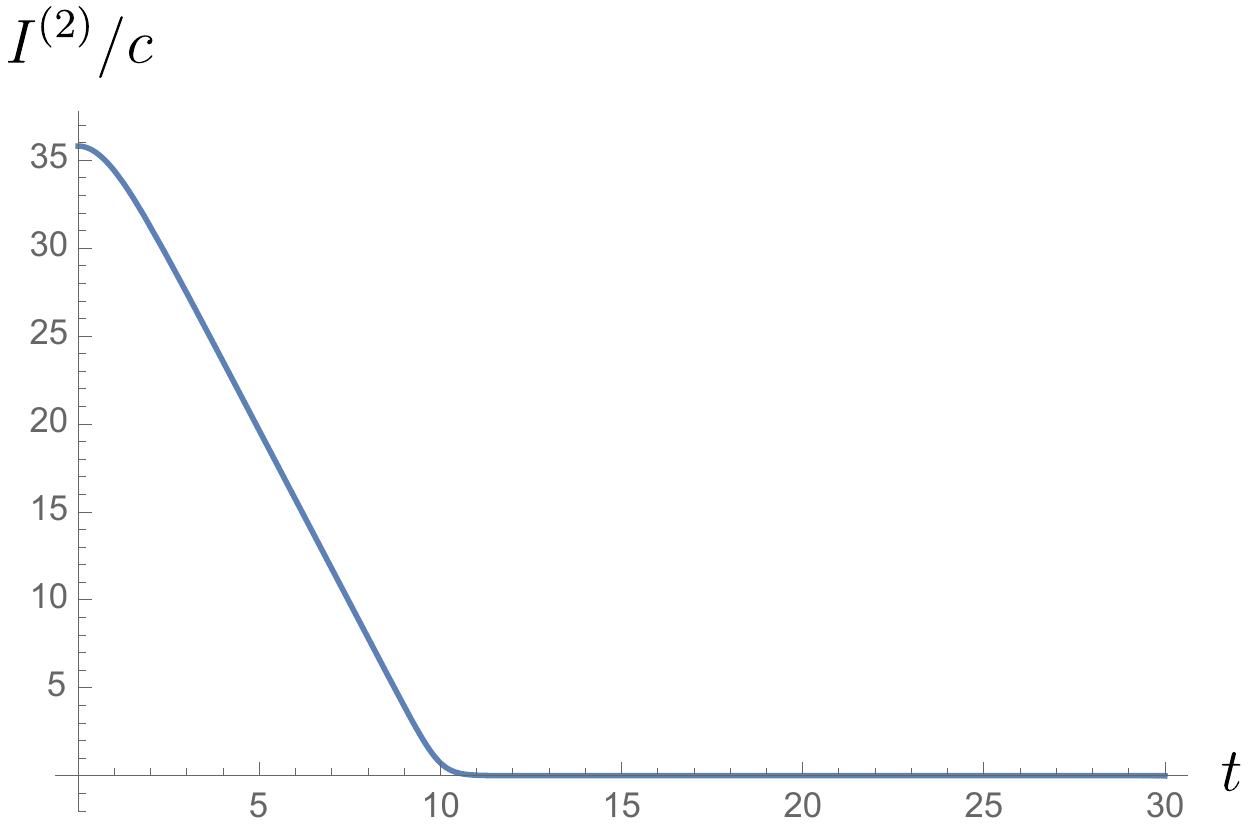}
    \includegraphics[height = 4.5cm]{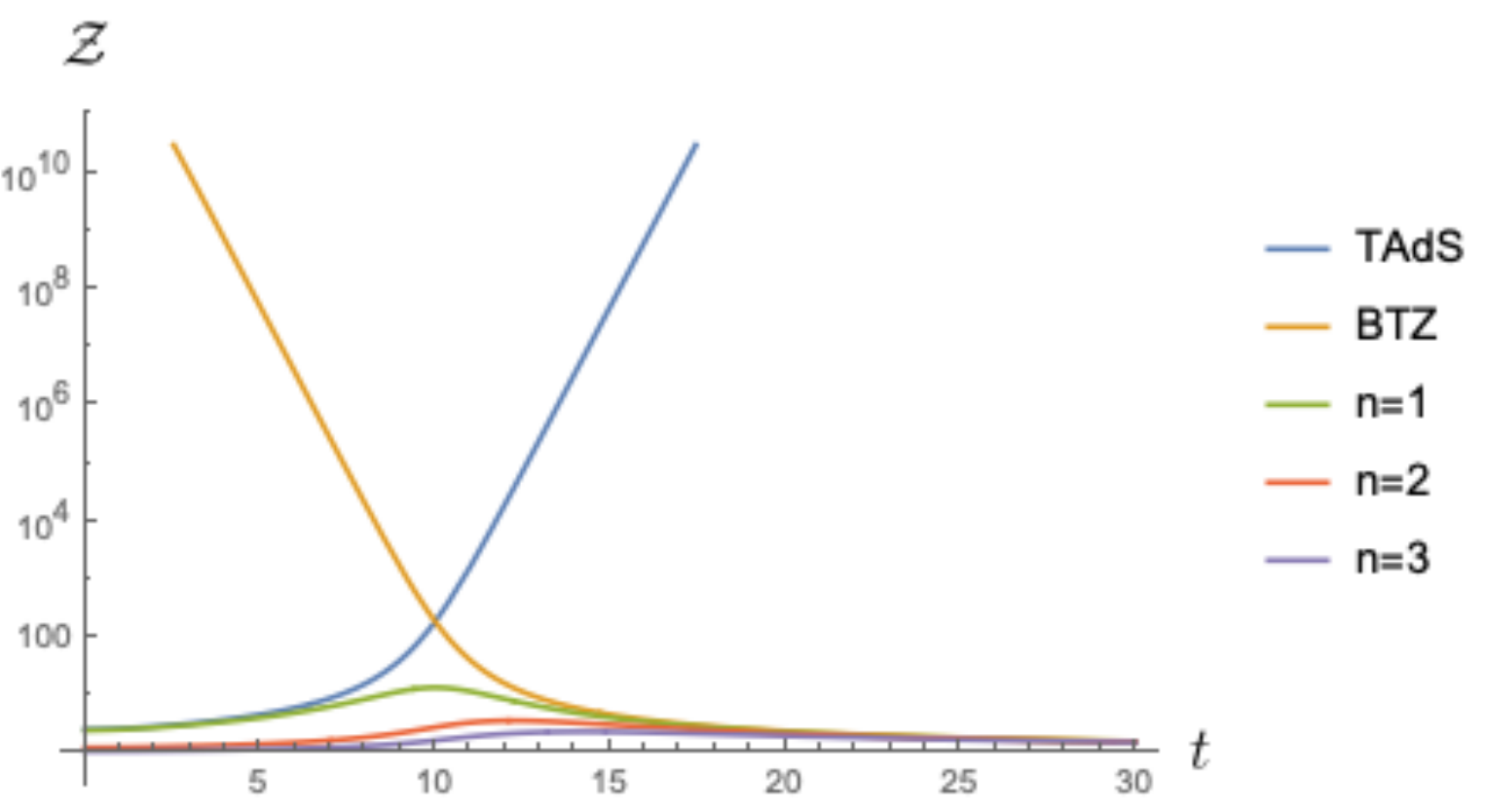}
    \caption{Left: the operator mutual information for symmetric intervals of length 10. Right: The corresponding contributions to the torus partition function. We see the Hawking-Page transition occur as the mutual information transitions for decreasing to constant (zero). In general, there will be a ``kink" in the operator mutual information that corresponds to the Hawking-Page transition. Note that all higher saddles are exponentially suppressed and never meaningfully contribute to the operator entanglement. }
    \label{holo_OMI_saddles}
\end{figure}

Operator mutual information was first computed for holographic CFTs in Ref.~\cite{2018arXiv181200013N}. The dynamical behavior strongly violates the quasi-particle picture and displays maximal scrambling. This was later explained in the context of an effective ``line-tension picture" \cite{2017PhRvX...7c1016N,2018arXiv180300089J,2018PhRvD..98j6025M,2018PhRvX...8b1013V} in Ref.~\cite{ 2019arXiv190607639K}. It is further illuminating to understand the operator entanglement for holographic theories and its phase transitions in terms of the torus partition function. Luckily, the cross-ratios \eqref{OMI_cross} are always real and between $0$ and $1$, so the only possible dominant saddles are the familiar thermal $AdS$ and BTZ black hole.
Interestingly, we find that the Thermal AdS partition function reproduces only the quasi-particle picture
\begin{align}
    I_3^{(TAdS)} = 0.
\end{align}
Therefore, the BTZ contribution is necessary for reproducing the maximal scrambling results of Ref.~\cite{2018arXiv181200013N}. In Fig.~\ref{holo_OMI_saddles}, we show the bipartite operator mutual information for symmetric intervals. The mutual information decays linearly until it hits zero where there is a sharp transition. In general, for bipartite mutual information in holography, there are sharp ``kinks" \cite{2018arXiv181200013N, 2019arXiv190607639K}. By studying the torus partition function, we have identified one of these entropic phase transitions with the Hawking-Page transition.

Having found the dominant contributions, let us compute the late-time value of TOMI. The bipartite operator entanglement is 
\begin{align}
    I^{(2)}(t) &= \log \left[\left| x\right|^{c/6}\left|2^8 (1-x) \right|^{-c/12} \left(\mathcal{Z}_{TAdS}(\tau, \bar{\tau})+\mathcal{Z}_{BTZ}(\tau, \bar{\tau})\right)\right] \nonumber
    \\
    % &= \log \left[\left| x\right|^{c/6}\left|2^8 (1-x) \right|^{-c/12} \right] + \frac{\pi i c}{12}\left(\bar{\tau} -\tau +\frac{1}{\tau} - \frac{1}{\bar{\tau}}\right)
    % \\
    &= \log \left[\left| x\right|^{c/6}\left|2^8 (1-x) \right|^{-c/12} \right] + \frac{\pi c}{12}\left(\frac{K(x)}{K(1-x)}+\frac{K(1-x)}{K(x)} + \frac{K(\bar{x})}{K(1-\bar{x})}+\frac{K(1-\bar{x})}{K(\bar{x})}\right).
\end{align}
As we did in Sec.~\ref{sec_RCFT}, for definiteness, let us take $A = (0,l), B_1 = (-\infty, 0), B_2 = (0,\infty)$. Again, the late-time behavior will be independent of this choice. The relevant cross-ratios are \eqref{crosses_OMI}.
The late time behavior for the semi-infinite subsystems are dominated by the thermal $AdS$ partition function, leading to information loss, while the full output system mutual information is dominated by the BTZ contribution and all information is retained. We consequently find
\begin{align}
    I^{(2)}_{AB_1}\rightarrow \frac{\pi c l}{24\epsilon}, \quad I^{(2)}_{AB_2}\rightarrow \frac{\pi c l}{24\epsilon},\quad I^{(2)}_{AB}\rightarrow \frac{\pi c l}{4\epsilon},\quad I_3^{(2)}\rightarrow -\frac{\pi c l}{6\epsilon}.
\end{align}
This is a surprising result that we need to unpack. First, note that $I_3^{(2)}$ is proportional to the subsystem size. This implies that an extensive amount of quantum information is scrambled by the holographic channel and is consistent with the results from Refs.~\cite{2018arXiv181200013N,2019arXiv190607639K}. More interestingly, we see that $I_3^{(2)}$
\textit{does not} saturate the fundamental bound from \eqref{I3_bound}. This is not an artifact of the bounds derived not being tight. For example, all R\'enyi TOMIs were found to saturate the lower bound for random unitary circuits in Ref.~\cite{2019arXiv190607639K}. The non-saturation is instead an artifact of $I_{AB_1}^{(2)}$ and $I_{AB_2}^{(2)}$ being positive at late times. This means that an extensive amount of information about region $A$ remains localized at late times. This is in tension with the common lore regarding the chaoticity of holographic conformal field theories and is the first example, to our knowledge, of holographic theories not appearing maximally chaotic.
\section{Discussion}

\label{sec_discussion}

There are many open questions that we have left for future work. One particular shortcoming of our work is that we have not understood the early-time chaos regime where the quantum Lyapunov exponent may be detected. It would be fascinating to understand systematically how the Lyapunov exponent and scrambling time emerge from the torus partition function. This could lead to further understanding of the chaos bound and theories which do not saturate the bound. More generally, it is of great interest to provide a more precise dictionary between the quantities. Some of the measures may prove to be redundant and others novel.

\acknowledgments

We would like to acknowledge insightful discussions with Nathan Benjamin, Yuya Kusuki, Chen-Te Ma, Masahiro Nozaki, and Mao Tian Tan. We are particularly grateful to Masahiro Nozaki for pointing out an error in a previous version of this paper.
S.R. is supported by
a Simons Investigator Grant from the Simons Foundation. L.N. is supported by the Kadanoff Fellowship at the University of Chicago.

% \bibliography{main}

\begin{thebibliography}{10}

\bibitem{PhysRevA.43.2046}
J.M.~Deutsch, {\it Quantum statistical mechanics in a closed system},  {\em
  Phys.~Rev.~A} {\bf 43} (Feb, 1991) 2046.

\bibitem{1994PhRvE..50..888S}
M.~{Srednicki}, {\it {Chaos and quantum thermalization}},  {\em
  Physical~Review~E} {\bf 50} (Aug, 1994) 888
  [\href{http://arxiv.org/abs/cond-mat/9403051}{{\tt cond-mat/9403051}}].

\bibitem{2014JHEP...03..067S}
S.H.~{Shenker} and D.~{Stanford}, {\it {Black holes and the butterfly effect}},
   {\em Journal~of~High~Energy~Physics} {\bf 3} (Mar., 2014) 67
  [\href{http://arxiv.org/abs/1306.0622}{{\tt arXiv:1306.0622}}].

\bibitem{2015PhRvL.115m1603R}
D.A.~{Roberts} and D.~{Stanford}, {\it {Diagnosing Chaos Using Four-Point
  Functions in Two-Dimensional Conformal Field Theory}},  {\em
  Physical~Review~Letters} {\bf 115} (Sept., 2015) 131603
  [\href{http://arxiv.org/abs/1412.5123}{{\tt arXiv:1412.5123}}].

\bibitem{2016JHEP...08..106M}
J.~{Maldacena}, S.H.~{Shenker} and D.~{Stanford}, {\it {A bound on chaos}},
  {\em Journal~of~High~Energy~Physics} {\bf 8} (Aug., 2016) 106
  [\href{http://arxiv.org/abs/1503.01409}{{\tt arXiv:1503.01409}}].

\bibitem{2019arXiv190901894B}
A.~{Bhattacharyya}, W.~{Chemissany}, S.~{Shajidul Haque} and B.~{Yan}, {\it
  {Towards the Web of Quantum Chaos Diagnostics}},  {\em arXiv~e-prints} (Sep,
  2019) arXiv:1909.01894 [\href{http://arxiv.org/abs/1909.01894}{{\tt
  arXiv:1909.01894}}].

\bibitem{2016AdPhy..65..239D}
L.~{D'Alessio}, Y.~{Kafri}, A.~{Polkovnikov} and M.~{Rigol}, {\it {From quantum
  chaos and eigenstate thermalization to statistical mechanics and
  thermodynamics}},  {\em Advances~in~Physics} {\bf 65} (May, 2016) 239
  [\href{http://arxiv.org/abs/1509.06411}{{\tt arXiv:1509.06411}}].

\bibitem{2017JHEP...11..048C}
J.~{Cotler}, N.~{Hunter-Jones}, J.~{Liu} and B.~{Yoshida}, {\it {Chaos,
  complexity, and random matrices}},  {\em Journal~of~High~Energy~Physics} {\bf
  2017} (Nov, 2017) 48 [\href{http://arxiv.org/abs/1706.05400}{{\tt
  arXiv:1706.05400}}].

\bibitem{2019PhLB..795..183D}
R.~{de Mello Koch}, J.H.~{Huang}, C.T.~{Ma} and H.J.R.~{Van Zyl}, {\it
  {Spectral form factor as an OTOC averaged over the Heisenberg group}},  {\em
  Physics~Letters~B} {\bf 795} (Aug, 2019) 183
  [\href{http://arxiv.org/abs/1905.10981}{{\tt arXiv:1905.10981}}].

\bibitem{2019arXiv190704289M}
C.T.~{Ma}, {\it {Early-Time and Late-Time Quantum Chaos}},  {\em
  arXiv~e-prints} (Jul, 2019) arXiv:1907.04289
  [\href{http://arxiv.org/abs/1907.04289}{{\tt arXiv:1907.04289}}].

\bibitem{2016JHEP...02..004H}
P.~{Hosur}, X.L.~{Qi}, D.A.~{Roberts} and B.~{Yoshida}, {\it {Chaos in quantum
  channels}},  {\em Journal~of~High~Energy~Physics} {\bf 2} (Feb., 2016) 4
  [\href{http://arxiv.org/abs/1511.04021}{{\tt arXiv:1511.04021}}].

\bibitem{2015JHEP...09..110A}
C.T.~{Asplund}, A.~{Bernamonti}, F.~{Galli} and T.~{Hartman}, {\it
  {Entanglement scrambling in 2d conformal field theory}},  {\em
  Journal~of~High~Energy~Physics} {\bf 2015} (Sep, 2015) 110
  [\href{http://arxiv.org/abs/1506.03772}{{\tt arXiv:1506.03772}}].

\bibitem{2019arXiv190302651Y}
B.~{Yan}, L.~{Cincio} and W.H.~{Zurek}, {\it {Information Scrambling and
  Loschmidt Echo}},  {\em arXiv~e-prints} (Mar, 2019) arXiv:1903.02651
  [\href{http://arxiv.org/abs/1903.02651}{{\tt arXiv:1903.02651}}].

\bibitem{2019JHEP...07..107R}
A.~{Romero-Berm{\'u}dez}, K.~{Schalm} and V.~{Scopelliti}, {\it {Regularization
  dependence of the OTOC. Which Lyapunov spectrum is the physical one?}},  {\em
  Journal~of~High~Energy~Physics} {\bf 2019} (Jul, 2019) 107
  [\href{http://arxiv.org/abs/1903.09595}{{\tt arXiv:1903.09595}}].

\bibitem{2018JHEP...07..002N}
Y.O.~{Nakagawa}, G.~{S{\'a}rosi} and T.~{Ugajin}, {\it {Chaos and relative
  entropy}},  {\em Journal~of~High~Energy~Physics} {\bf 2018} (Jul, 2018) 2
  [\href{http://arxiv.org/abs/1805.01051}{{\tt arXiv:1805.01051}}].

\bibitem{2017JHEP...04..121R}
D.A.~{Roberts} and B.~{Yoshida}, {\it {Chaos and complexity by design}},  {\em
  Journal~of~High~Energy~Physics} {\bf 2017} (Apr, 2017) 121
  [\href{http://arxiv.org/abs/1610.04903}{{\tt arXiv:1610.04903}}].

\bibitem{2018arXiv181208657P}
D.E.~{Parker}, X.~{Cao}, A.~{Avdoshkin}, T.~{Scaffidi} and E.~{Altman}, {\it {A
  Universal Operator Growth Hypothesis}},  {\em arXiv~e-prints} (Dec, 2018)
  arXiv:1812.08657 [\href{http://arxiv.org/abs/1812.08657}{{\tt
  arXiv:1812.08657}}].

\bibitem{2019arXiv190610808M}
C.~{Murthy} and M.~{Srednicki}, {\it {Bounds on chaos from the eigenstate
  thermalization hypothesis}},  {\em arXiv~e-prints} (Jun, 2019)
  arXiv:1906.10808 [\href{http://arxiv.org/abs/1906.10808}{{\tt
  arXiv:1906.10808}}].

\bibitem{2020arXiv200514243K}
J.~{Kudler-Flam}, M.~{Nozaki}, S.~{Ryu} and M.~{Tian Tan}, {\it {Entanglement
  of Local Operators and the Butterfly Effect}},  {\em arXiv~e-prints} (May,
  2020) arXiv:2005.14243 [\href{http://arxiv.org/abs/2005.14243}{{\tt
  arXiv:2005.14243}}].

\bibitem{2017JHEP...05..118C}
J.S.~{Cotler}, G.~{Gur-Ari}, M.~{Hanada}, J.~{Polchinski}, P.~{Saad},
  S.H.~{Shenker}, D.~{Stanford}, A.~{Streicher} and M.~{Tezuka}, {\it {Black
  holes and random matrices}},  {\em Journal~of~High~Energy~Physics} {\bf 5}
  (May, 2017) 118 [\href{http://arxiv.org/abs/1611.04650}{{\tt
  arXiv:1611.04650}}].

\bibitem{2017JHEP...03..154B}
V.~{Balasubramanian}, B.~{Craps}, B.~{Czech} and G.~{S{\'a}rosi}, {\it {Echoes
  of chaos from string theory black holes}},  {\em
  Journal~of~High~Energy~Physics} {\bf 2017} (Mar, 2017) 154
  [\href{http://arxiv.org/abs/1612.04334}{{\tt arXiv:1612.04334}}].

\bibitem{1969JETP...28.1200L}
A.I.~{Larkin} and Y.N.~{Ovchinnikov}, {\it {Quasiclassical Method in the Theory
  of Superconductivity}},  {\em
  Soviet~Journal~of~Experimental~and~Theoretical~Physics} {\bf 28} (June, 1969)
  1200.

\bibitem{2001PhRvA..63d0304Z}
P.~{Zanardi}, {\it {Entanglement of quantum evolutions}},  {\em
  Physical~Review~A} {\bf 63} (Apr, 2001) 040304
  [\href{http://arxiv.org/abs/quant-ph/0010074}{{\tt quant-ph/0010074}}].

\bibitem{2007PhRvA..76c2316P}
T.~{Prosen} and I.~{Pi{\v{z}}orn}, {\it {Operator space entanglement entropy in
  a transverse Ising chain}},  {\em Physical~Review~A} {\bf 76} (Sep, 2007)
  032316 [\href{http://arxiv.org/abs/0706.2480}{{\tt arXiv:0706.2480}}].

\bibitem{2009PhRvB..79r4416P}
I.~{Pi{\v{z}}orn} and T.~{Prosen}, {\it {Operator space entanglement entropy in
  XY spin chains}},  {\em Physical~Review~B} {\bf 79} (May, 2009) 184416
  [\href{http://arxiv.org/abs/0903.2432}{{\tt arXiv:0903.2432}}].

\bibitem{2017JPhA...50w4001D}
J.~{Dubail}, {\it {Entanglement scaling of operators: a conformal field theory
  approach, with a glimpse of simulability of long-time dynamics in
  1{\,}{\,}+{\,}{\,}1d}},  {\em Journal~of~Physics~A~Mathematical~General} {\bf
  50} (Jun, 2017) 234001 [\href{http://arxiv.org/abs/1612.08630}{{\tt
  arXiv:1612.08630}}].

\bibitem{2017PhRvB..95i4206Z}
T.~{Zhou} and D.J.~{Luitz}, {\it {Operator entanglement entropy of the time
  evolution operator in chaotic systems}},  {\em Physical~Review~B} {\bf 95}
  (Mar, 2017) 094206 [\href{http://arxiv.org/abs/1612.07327}{{\tt
  arXiv:1612.07327}}].

\bibitem{2018arXiv180300089J}
C.~{Jonay}, D.A.~{Huse} and A.~{Nahum}, {\it {Coarse-grained dynamics of
  operator and state entanglement}},  {\em arXiv~e-prints} (Feb., 2018)
  [\href{http://arxiv.org/abs/1803.00089}{{\tt arXiv:1803.00089}}].

\bibitem{2018arXiv181200013N}
L.~{Nie}, M.~{Nozaki}, S.~{Ryu} and M.~{Tian Tan}, {\it {Signature of quantum
  chaos in operator entanglement in 2d CFTs}},  {\em arXiv~e-prints} (Nov,
  2018) arXiv:1812.00013 [\href{http://arxiv.org/abs/1812.00013}{{\tt
  arXiv:1812.00013}}].

\bibitem{2019arXiv190607639K}
J.~{Kudler-Flam}, M.~{Nozaki}, S.~{Ryu} and M.T.~{Tan}, {\it {Quantum vs.
  classical information: operator negativity as a probe of scrambling}},  {\em
  Journal~of~High~Energy~Physics} {\bf 2020} (Jan., 2020) 31
  [\href{http://arxiv.org/abs/1906.07639}{{\tt arXiv:1906.07639}}].

\bibitem{CHOI1975285}
M.D.~Choi, {\it Completely positive linear maps on complex matrices},  {\em
  Linear~Algebra~and~its~Applications} {\bf 10} (1975) 285 .

\bibitem{JAMIOLKOWSKI1972275}
A.~Jamiolkowski, {\it Linear transformations which preserve trace and positive
  semidefiniteness of operators},  {\em Reports~on~Mathematical~Physics} {\bf
  3} (1972) 275 .

\bibitem{2005JSMTE..04..010C}
P.~{Calabrese} and J.~{Cardy}, {\it {Evolution of entanglement entropy in
  one-dimensional systems}},  {\em
  Journal~of~Statistical~Mechanics:~Theory~and~Experiment} {\bf 2005} (Apr,
  2005) 04010 [\href{http://arxiv.org/abs/cond-mat/0503393}{{\tt
  cond-mat/0503393}}].

\bibitem{2006PhRvL..96m6801C}
P.~{Calabrese} and J.~{Cardy}, {\it {Time Dependence of Correlation Functions
  Following a Quantum Quench}},  {\em Physical~Review~Letters} {\bf 96} (Apr,
  2006) 136801 [\href{http://arxiv.org/abs/cond-mat/0601225}{{\tt
  cond-mat/0601225}}].

\bibitem{2014PhRvL.112k1602N}
M.~{Nozaki}, T.~{Numasawa} and T.~{Takayanagi}, {\it {Quantum Entanglement of
  Local Operators in Conformal Field Theories}},  {\em Physical~Review~Letters}
  {\bf 112} (Mar, 2014) 111602 [\href{http://arxiv.org/abs/1401.0539}{{\tt
  arXiv:1401.0539}}].

\bibitem{2017PhRvD..96d6020C}
P.~{Caputa}, Y.~{Kusuki}, T.~{Takayanagi} and K.~{Watanabe}, {\it
  {Out-of-time-ordered correlators in a (T$^{2}$)$^{n}$/Z$_{n}$ CFT}},  {\em
  Physical~Review~D} {\bf 96} (Aug, 2017) 046020
  [\href{http://arxiv.org/abs/1703.09939}{{\tt arXiv:1703.09939}}].

\bibitem{2001CMaPh.219..399L}
O.~{Lunin} and S.D.~{Mathur}, {\it {Correlation Functions for M$^{N}$/S$_{N}$
  Orbifolds}},  {\em Communications~in~Mathematical~Physics} {\bf 219} (2001)
  399 [\href{http://arxiv.org/abs/hep-th/0006196}{{\tt hep-th/0006196}}].

\bibitem{2019JHEP...04..025B}
N.~{Benjamin}, E.~{Dyer}, A.L.~{Fitzpatrick} and Y.~{Xin}, {\it {The most
  irrational rational theories}},  {\em Journal~of~High~Energy~Physics} {\bf
  2019} (Apr, 2019) 25 [\href{http://arxiv.org/abs/1812.07579}{{\tt
  arXiv:1812.07579}}].

\bibitem{2016JHEP...08..129G}
Y.~{Gu} and X.L.~{Qi}, {\it {Fractional statistics and the butterfly effect}},
  {\em Journal~of~High~Energy~Physics} {\bf 8} (Aug., 2016) 129
  [\href{http://arxiv.org/abs/1602.06543}{{\tt arXiv:1602.06543}}].

\bibitem{2016PTEP.2016k3B06C}
P.~{Caputa}, T.~{Numasawa} and A.~{Veliz-Osorio}, {\it {Out-of-time-ordered
  correlators and purity in rational conformal field theories}},  {\em
  Progress~of~Theoretical~and~Experimental~Physics} {\bf 2016} (Nov, 2016)
  113B06 [\href{http://arxiv.org/abs/1602.06542}{{\tt arXiv:1602.06542}}].

\bibitem{2019arXiv190502191K}
Y.~{Kusuki} and M.~{Miyaji}, {\it {Entanglement Entropy, OTOC and Bootstrap in
  2D CFTs from Regge and Light Cone Limits of Multi-point Conformal Block}},
  {\em arXiv~e-prints} (May, 2019) arXiv:1905.02191
  [\href{http://arxiv.org/abs/1905.02191}{{\tt arXiv:1905.02191}}].

\bibitem{2017JHEP...08..075D}
E.~{Dyer} and G.~{Gur-Ari}, {\it {2D CFT partition functions at late times}},
  {\em Journal~of~High~Energy~Physics} {\bf 2017} (Aug, 2017) 75
  [\href{http://arxiv.org/abs/1611.04592}{{\tt arXiv:1611.04592}}].

\bibitem{DiFrancesco:1997nk}
P.~Di~Francesco, P.~Mathieu and D.~Senechal, {\it {Conformal Field Theory}},
  Graduate Texts in Contemporary Physics Springer-Verlag, New York (1997).
%%CITATION = INSPIRE-454643;%%

\bibitem{Belavin:1984vu}
A.A.~Belavin, A.M.~Polyakov and A.B.~Zamolodchikov, {\it {Infinite Conformal
  Symmetry in Two-Dimensional Quantum Field Theory}},  {\em Nucl.~Phys.} {\bf
  B241} (1984) 333. [,605(1984)].
%%CITATION = NUPHA,B241,333;%%

\bibitem{2001JHEP...09..006R}
I.~{Runkel} and G.M.T.~{Watts}, {\it {A non-rational CFT with c = 1 as a limit
  of minimal models}},  {\em Journal~of~High~Energy~Physics} {\bf 9} (Sept.,
  2001) 006 [\href{http://arxiv.org/abs/hep-th/0107118}{{\tt hep-th/0107118}}].

\bibitem{bezout2010theorie}
E.~Bezout, {\it Theorie generale des equations algebriques (1779)},  Kessinger
  Publishing (2010).

\bibitem{Chen2017}
H.~Chen, C.~Hussong, J.~Kaplan and D.~Li, {\it A numerical approach to virasoro
  blocks and the information paradox},  {\em Journal~of~High~Energy~Physics}
  {\bf 2017} (Sep, 2017) 102.

\bibitem{1999IJTP...38.1113M}
J.~{Maldacena}, {\it {The Large-N Limit of Superconformal Field Theories and
  Supergravity}},  {\em International~Journal~of~Theoretical~Physics} {\bf 38}
  (Jan, 1999) 1113 [\href{http://arxiv.org/abs/hep-th/9711200}{{\tt
  hep-th/9711200}}].

\bibitem{2014JHEP...09..118H}
T.~{Hartman}, C.A.~{Keller} and B.~{Stoica}, {\it {Universal spectrum of 2d
  conformal field theory in the large c limit}},  {\em
  Journal~of~High~Energy~Physics} {\bf 2014} (Sep, 2014) 118
  [\href{http://arxiv.org/abs/1405.5137}{{\tt arXiv:1405.5137}}].

\bibitem{2008JHEP...10..065S}
Y.~{Sekino} and L.~{Susskind}, {\it {Fast scramblers}},  {\em
  Journal~of~High~Energy~Physics} {\bf 2008} (Oct, 2008) 065
  [\href{http://arxiv.org/abs/0808.2096}{{\tt arXiv:0808.2096}}].

\bibitem{Hawking1983}
S.W.~Hawking and D.N.~Page, {\it Thermodynamics of black holes in anti-de
  sitter space},  {\em Communications~in~Mathematical~Physics} {\bf 87} (Dec,
  1983) 577.

\bibitem{1992PhRvL..69.1849B}
M.~{Banados}, C.~{Teitelboim} and J.~{Zanelli}, {\it {Black hole in
  three-dimensional spacetime}},  {\em Physical~Review~Letters} {\bf 69} (Sep,
  1992) 1849 [\href{http://arxiv.org/abs/hep-th/9204099}{{\tt
  hep-th/9204099}}].

\bibitem{2010JHEP...02..029M}
A.~{Maloney} and E.~{Witten}, {\it {Quantum gravity partition functions in
  three dimensions}},  {\em Journal~of~High~Energy~Physics} {\bf 2010} (Feb,
  2010) 29 [\href{http://arxiv.org/abs/0712.0155}{{\tt arXiv:0712.0155}}].

\bibitem{2015JHEP...02..080K}
C.A.~{Keller} and A.~{Maloney}, {\it {Poincar{\'e} series, 3D gravity and CFT
  spectroscopy}},  {\em Journal~of~High~Energy~Physics} {\bf 2015} (Feb, 2015)
  80 [\href{http://arxiv.org/abs/1407.6008}{{\tt arXiv:1407.6008}}].

\bibitem{2003JHEP...04..021M}
J.~{Maldacena}, {\it {Eternal black holes in anti-de Sitter}},  {\em
  Journal~of~High~Energy~Physics} {\bf 2003} (Apr, 2003) 021
  [\href{http://arxiv.org/abs/hep-th/0106112}{{\tt hep-th/0106112}}].

\bibitem{2017PhRvX...7c1016N}
A.~{Nahum}, J.~{Ruhman}, S.~{Vijay} and J.~{Haah}, {\it {Quantum Entanglement
  Growth under Random Unitary Dynamics}},  {\em Physical~Review~X} {\bf 7}
  (July, 2017) 031016 [\href{http://arxiv.org/abs/1608.06950}{{\tt
  arXiv:1608.06950}}].

\bibitem{2018PhRvD..98j6025M}
M.~{Mezei}, {\it {Membrane theory of entanglement dynamics from holography}},
  {\em Physical~Review~D} {\bf 98} (Nov., 2018) 106025
  [\href{http://arxiv.org/abs/1803.10244}{{\tt arXiv:1803.10244}}].

\bibitem{2018PhRvX...8b1013V}
C.W.~{von Keyserlingk}, T.~{Rakovszky}, F.~{Pollmann} and S.L.~{Sondhi}, {\it
  {Operator Hydrodynamics, OTOCs, and Entanglement Growth in Systems without
  Conservation Laws}},  {\em Physical~Review~X} {\bf 8} (Apr., 2018) 021013
  [\href{http://arxiv.org/abs/1705.08910}{{\tt arXiv:1705.08910}}].

\end{thebibliography}
% \bibliographystyle{JHEP}

\providecommand{\href}[2]{#2}\begingroup\raggedright\endgroup

\end{document}